\documentclass[superscriptaddress, prd, aps,amsmath,amssymb,showpacs,showkeys, onecolumn]{revtex4-2}
\usepackage[dvips]{graphicx,color}
\usepackage{subfig}
\usepackage{times}
\usepackage{xcolor}
\usepackage{graphicx, amsmath}
\usepackage{float}
\usepackage{multirow}

\usepackage{booktabs}

\usepackage{caption}
\usepackage{subcaption}
\usepackage[%
  colorlinks=true,
  urlcolor=blue,
  linkcolor=red,
  citecolor=blue
]{hyperref}

\usepackage{mathrsfs}
\usepackage{orcidlink}
\DeclareMathAlphabet{\pazocal}{OMS}{zplm}{m}{n}

\newcommand{\Lb}{\pazocal{L}}

\begin{document}
\title{Geodesic structure, eikonal quasinormal modes and thermodynamic properties of a Schwarzschild-de Sitter-like black hole with global monopole in Bumblebee gravity}

\author{Irengbam Roshila Devi}
\email[Email: ]{roshilairengbam@gmail.com}
\affiliation{Department of Mathematics, Manipur University, Canchipur,
Imphal, Manipur, India-795003}

\author{Yenshembam Priyobarta Singh}
\email[Email: ]{priyoyensh@gmail.com}
\affiliation{Department of Mathematics, Manipur University, Canchipur,
Imphal, Manipur, India-795003}

\author{Dhruba Jyoti Gogoi\orcidlink{0000-0002-4776-8506}}
\email[Email: ]{moloydhruba@yahoo.in}
\affiliation{Department of Physics, Madhabdev University, Narayanpur, Lakhimpur 784164, Assam, India}
\affiliation{Research Center of Astrophysics and Cosmology, Khazar University, Baku, AZ1096, 41 Mehseti Street, Azerbaijan}

\author{Telem Ibungochouba Singh}
\email[Email: ]{ibungochouba@rediffmail.com}
\affiliation{Department of Mathematics, Manipur University, Canchipur,
Imphal, Manipur, India-795003}

%\date{}
\begin{abstract}

We investigate the geodesic dynamics, eikonal quasinormal modes (QNMs), and extended phase space thermodynamics of a Schwarzschild-de Sitter-like black hole coupled with a global monopole in Bumblebee gravity. Our analysis reveals that spontaneous Lorentz symmetry breaking and topological defects modify the effective radial dynamics, leading to lower effective potential barriers and a reduced magnitude of the effective force. For null geodesics, higher monopole parameters expand the photon sphere and critical impact parameter, implying an enlarged black hole shadow, while mitigating dynamical instability. For massive particles, the stable circular orbit window narrows significantly as the Outermost Stable Circular Orbit (OSCO) shrinks, alongside a marked enhancement in perihelion precession. In addition, we examine the correspondence between the eikonal QNMs and the properties of null geodesics. The excellent agreement between the eikonal approximation and the WKB method for scalar and electromagnetic perturbations supports the validity of the correspondence for the spacetime under consideration. In the extended thermodynamic phase space, we evaluate the black hole as a holographic heat engine. We demonstrate that classical efficiency is mathematically blind to Lorentz violation; however, introducing a quantum-corrected modified entropy tightly couples the macroscopic work output to the symmetry-breaking framework. Ultimately, the necessity to respect the Carnot bound imposes a strict macroscopic constraint on the Lorentz-violating parameter, safeguarding the Second Law of Thermodynamics.
\end{abstract}
\maketitle

\section{Introduction}
Black holes (BHs) are intriguing objects in our universe that have gained extensive interest, ever since their existence was predicted by Einstein in General Relativity. In recent years, GR has received unprecedented observational validation through the monumental detection of Gravitational Waves (GWs) \cite{B.P. Abbott 2016a, LIGOScientific:2017ync, 2,3,4,5}. Also, the observation of the shadow of the BH in the M87$^{*}$ galaxy by the Event Horizon Telescope (EHT) collaboration has sparked increased interest in BH physics in recent years \cite{K. Akiyama L1 2019, K. Akiyama L5 2019, K. Akiyama L6 2019, Roshila 2026}. GR encounters significant challenges despite its many successes, like it's inability to incorporate quantum mechanics, which describes physical phenomena at the microscopic level \cite{Faizuddin2026}. This incompatibility continues to hinder the pursuit for a viable quantum gravity (QG) framework. Alternative theories like string theory and Loop Quantum Gravity incorporate quantum modifications to classical spacetime, particularly in regions close to singularities and black hole horizons \cite{L. Modesto 2010, J. Alfaro 2002}. The Standard Model of particle physics has been effective in explaining the fundamental particles and their quantum-scale interactions. Significant uncertainties in determining the fundamental parameters of black holes precisely leave room for the development of several theories of modified gravity \cite{R. Konoplya 2016}. The Bumblebee gravity framework is one of the modified theories that will be especially relevant, as it incorporates Lorentz symmetry violation (LSV) through a non-zero vacuum expectation value of the Bumblebee field when an appropriate potential is considered \cite{YenshembamNuclear2025, Jayasri 2026, Roshila 2026}. One of the fundamental principles of modern physics is Lorentz symmetry, which asserts that the laws of physics remain invariant in all inertial reference frames. The study of Lorentz symmetry breaking (LSB) offers valuable insights into the fundamental principles of physics and the underlying nature of spacetime. An exact Schwarzschild-like BH solution was found by Casana et al. based on the Bumblebee gravity framework \cite{R. Casana 2018}. Numerous exact BH solutions have been derived by researchers within this framework since then \cite{R. V. Maluf 2021a,J.Z. Liu 2025,C. Ding 2020,I. Gullu 2022,C. Ding 2022}. Numerous properties of black holes have been explored within this framework such as Hawking radiation \cite{S. Kanzi 2019}, thermodynamic aspects \cite{R. Karmakar 2023,Z.F. Mai 2023,Y.S. An 2024}, quasinormal modes \cite{X. Zhang 2023,D.J. Gogoi 2022}, BH shadow \cite{R. V. Maluf 2021a, S.K. Jha 2021}, gravitational lensing \cite{A. Ovgun 2018}, etc.

At the same time, topological spacetime defects such as global monopole have drawn substantial interest because of their cosmological relevance and their effects on the structure of BH spacetimes \cite{Ahmad2025}. According to Grand Unified Theories, global monopoles are a particular type of topological defect which may have been formed during the early stage of the universe via the spontaneous breaking of the global O(3) symmetry into U(1) \cite{Mengjie 2021,T.W.B. 1976,A. Vilenkin 1985}. The spacetime geometry is altered by these defects owing to their energy contributions, giving rise to a solid angle deficit that influences geodesic motion and other physical phenomena \cite{Ahmad2025}. In the context of Bumblebee gravity, which incorporates spontaneous LSB, spherically symmetric BH solutions containing a global monopole (GM) have been studied to explore the combined effects of topological defects and Bumblebee vector field dynamics \cite{Z. Malik 2024}.

Over the past few years, investigations of the dynamics of test particles have played a significant role in testing gravitational theories understanding astrophysical phenomena, providing valuable insights about the gravitational field in the proximity of BHs \cite{MouXu2025}. The structure of geodesics of timelike particles and photons in the proximity of BHs is strongly affected by the nature of the geometry of the spacetime \cite{MouXu2025}. The study of the geodesic structure offers essential insights about the geometric and causal nature of a spacetime, describing the motion of  massive particles and light rays in the curved manifold \cite{Ahmad2025}. Null geodesics are essential for investigating phenomena including gravitational lensing effects, photon spheres, BH shadow, etc., while timelike geodesics or the trajectories of massive particles characterize the orbital motion and associated precession phenomena \cite{V. Cardoso 2009,F. Ahmed 2025,J. C. Drawer,Y. C. Ko 2024,K. Chen 2024}. These geodesic structures often manifest characteristic features that depart from the predictions of GR in modified gravity theories, thereby providing observational tests for alternative theories of gravity \cite{Ahmad2025}. The stability characteristics of circular orbits and the strength of the gravitational binding acting upon the test particles can be characterized by the investigation of effective potentials, impact parameters, Lyapunov exponents, etc., in the modified spacetimes. In particular, the presence of the LSB, GM parameter and the cosmological constant may substantially modify the BH dynamics, thereby altering the surrounding spacetime structure of the BH.

 Motivated by the above discussion, we investigate the null geodesic and timelike structures of a Schwarzchild-like BH with a cosmological constant and GM within the Bumblebee gravity to gain better insight into the surrounding spacetime geometry. Refs. \cite{D.J. Gogoi 2022,YenshembamNuclear2025} have already derived and discussed the exterior spacetime geometry of this black hole. Although the spacetime was used in our previous work \cite{YenshembamNuclear2025} to investigate the different types of perturbations and their associated quasinormal modes, the analysis of null and timelke geodesics and their corresponding trajectories as well as the eikonal quasinormal modes was not addressed. In this study, we probe how the Lorentz violation (LV) parameter and 
 the GM parameter alter the spacetime structure, affect the geodesic motion, stability of circular orbits, the innermost stable circular orbit (ISCO), the outermost circular orbit (OSCO) and the characteristic modes in the eikonal limit. We examine how these parameters alter physical observables, like the effective potential which governs the motion of test particles, trajectories of photons and massive particles, dynamics of particles around the black hole, etc., 
 
 Among the numerous techniques used to investigate the geodesic stability, the Lyapunov exponent has been widely employed \cite{Shobhit 2022, Roshila 2026}. The Lyapunov exponent is the mean rate of seperation of two nearby trajectories \cite{Shobhit 2022, Roshila 2026}. The Lyapunov exponent is negative or positive corresponding to the convergence or divergence of two nearby geodesics \cite{V. Cardoso 2009,M. Sharif 2017, Shobhit 2022}.
 
 Quasinormal modes (QNMs) constitute one of the most remarkable characteristics of BHs which can be detected with gravitational-wave interferometers. BH spacetimes are inherently dissipative owing to the presence of an event horizon, causing quasinormal frequencies to be complex in general \cite{Mengjie 2021}. QNMs are distinctive features of BHs which describe the damped oscillations of spacetime caused by external perturbations and the QNM spectra are influenced by the parameters of BHs like spin and mass \cite{Kokkotas 1999, Jose 2023, Yating 2025}. The ringdown phase of gravitational-wave emission is characterized by QNMs and they encode information about the geometry of the surrounding spacetime \cite{Faizuddin2026}. Groundbreaking observations by the VIRGO and LIGO collaborations have firmly established the existence of these signals \cite{B.P. Abbott 2016a, B.P. Abbott 2016b, 2}. Consequently, they serve as an essential tool for exploring the characteristics of black holes. The quasinormal mode frequencies are complex vibrations where the real part corresponds to the oscillation frequency and the imaginary part indicates the decay rate of the amplitude of the oscillation \cite{Jose 2023, R.A. Konoplya 2011}. There are numerous techniques to compute QNMs like Wentzel-Kramers-Brillouin (WKB) method \cite{R.A. Konoplya 2011, R.A. Konoplya 2019, N Media 2025, Y.S. Priyobarta 2024, Y.S. Priyobarta 2025c}, the Asymptotic Iteration Method (AIM) \cite{Y.S. Priyobarta 2025b, Roshila 2026}, the time-domain integration method, the continued fraction method, etc. 
 
In this paper, we will compute the QNMs by employing an alternative well-known method, which utilizes Lyapunov exponents \cite{Dias 2018, Cornish 2003, Vitor 2009} in addition to the WKB method. In the eikonal (geometric optics) limit, the quasinormal frequencies, also referred to as photon-sphere QNMs, are governed by the characteristics of unstable circular photon orbits in the equatorial plane \cite{Alexander 2022} . Eikonal QNMs $(\ell \gg 1)$ constitute a class of oscillations that are especially useful for exploring the high-frequency behaviour of QNMs. The real part of the frequency is closely connected to the unstable circular photon orbits of the BH, in the eikonal limit \cite{Jose 2023}. Ferrari and Mashhoon \cite{V. Ferrari 1984} first proposed the correspondence between Lyapunov exponents and QNMs. It was subsequently extended to spacetimes which are asymptotically flat, stationary and spherically symmetric \cite{Vitor 2009, I.Z. Stefanov 2010,M.D. Johnson 2020}. For a Schwarzschild BH in the eikonal limit, it was shown that the real part of the QNM frequency is associated with the angular velocity of the circular null orbit, whereas the imaginary part is associated with the Lyapunov exponent, which quantifies the orbital instability timescale. This connection was further employed to determine the QNMs of the dual stringy black hole \cite{S. Giri 2021}, as well as the noncommutative Schwarzschild black hole \cite{S. Giri 2022}. In this paper, the fundamental overtone QNMs will be considered for large overtone numbers in the spacetime. The WKB method will be used for computing the QNMs, and the correspondence between the characteristics of the null geodesics of the BH and the QNMs will be investigated.

BH thermodynamics has emerged as one of the major areas of research in theoretical physics, since the pioneering contributions of J. Bekenstein and S. Hawking to the thermodynamics of black holes \cite{J.D. Bekenstein 1973, S.W. Hawking 1975, Dhruba 2026}. More specifically, it has evolved into a fundamental interface connecting gravitation, statistical physics and quantum theory \cite{ Dhruba 2026}. A comprehensive understanding of BHs requires the inclusion of quantum effects, underscoring the outstanding quest for a consistent theory of quantum gravity, resolving related cosmological problems and unifying the fundamental interactions \cite{S. Rani 2024, Dhruba 2026}. Extended phase space thermodynamics, in which the cosmological constant is identified with thermodynamic pressure, provides a framework for describing BHs as the working substance of macroscopic heat engines \cite{Dhruba 2026}. This enables BHs to execute cyclic processes like Stirling, Carnot and Brayton cycles, thereby producing net mechanical work \cite{Dhruba 2026}.
 
The layout of this paper is as follows: In Sect. 2, we briefly introduce the spacetime, and the effect of the GM and LV parameters on the two horizons. In Sect. 3, we examine the geodesics within the Schwarzschild-dS-like BH spacetime and find the expression for the effective potentials. Section 4 examines the null geodesics in the spacetime, including the effective force, photon spheres, null trajectories and Lyapunov exponents. In Sect. 5, we analyze the timelike potential, massive circular orbits, massive particle trajectories, precession of the orbits and stability. In Sect. 6, the eikonal QNMs are discussed, and their correspondence with the null geodesics is examined. In Sect. 7, the black hole is evaluated as a holographic heat engine in the extended thermodynamic phase space. Finally, in Sect. 8, we present our conclusion.

\section{Static spherically symmetric BH with a cosmological constant and GM in Bumblebee gravity}
The metric of a Schwarzschild-dS-like BH with a cosmological constant and GM can be expressed as \cite{YenshembamNuclear2025}
\begin{eqnarray}\label{eqn line element}
ds^2=f(r) dt^2 -\dfrac{1+L}{f(r)}dr^2-r^2 d\theta^2-r^2 \sin^2\theta d \varphi^2,
\end{eqnarray}
where the lapse function $f(r)$ is expressed as 
\begin{eqnarray}
f(r)=1-\kappa \eta^2-\dfrac{2 M}{r}-\dfrac{(1+L)\Lambda r^2}{3}.
\end{eqnarray}
Here, $L$, $\eta$, and $\Lambda$ represents the LV parameter, GM parameter and the cosmological constant. M is the mass of the black hole and $\kappa=\dfrac{8 \pi G}{c^4}=8 \pi$, since $G=1$ and $c=1$ in the natural units. When $\eta=0$ and $\Lambda=0$, Eq. \eqref{eqn line element} degenerates to the Schwarzschild-dS-like BH \cite{R. V. Maluf 2021} and the Schwarzschild-like solution with GM \cite{Ibrahim 2022}, respectively. Finally, in the limit $L \rightarrow 0$, the Schwarzschild-dS BH with GM is recovered \cite{YenshembamNuclear2025}. The metric \eqref{eqn line element} features the event horizon $r_h$ and the cosmological horizon $r_c$ \cite{YenshembamNuclear2025}:
\begin{eqnarray}
r_h=2\sqrt{\dfrac{1-\kappa \eta^2}{(1+L)\Lambda}}\cos \bigg(\dfrac{\Phi+\pi}{3}\bigg),\hspace{0.6 cm}
r_c=2\sqrt{\dfrac{1-\kappa \eta^2}{(1+L)\Lambda}}\cos \bigg(\dfrac{ -\Phi+ \pi}{3}\bigg),
\end{eqnarray}
where $ \Phi =\cos^{-1}\left[\sqrt{\dfrac{9 M^2 (1+L)\Lambda}{(1-\kappa \eta ^{2})^3} }\right]  $ . 
The condition $ 9(1+L)M^2 \Lambda < \left(1-\kappa \eta^2 \right)^3$ must be satisfied for the existence of the two horizons.

\begin{figure}[h!]
\centering
  \subfloat[\centering ]{{\includegraphics[width=170pt,height=160pt]{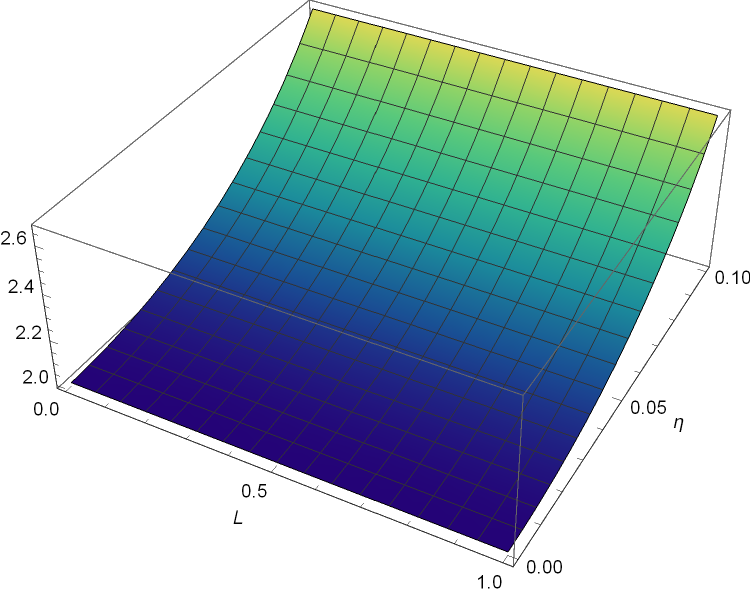}}\label{fig eventhorizon}}
  \qquad
   \subfloat[\centering ]{{\includegraphics[width=170pt,height=160pt]{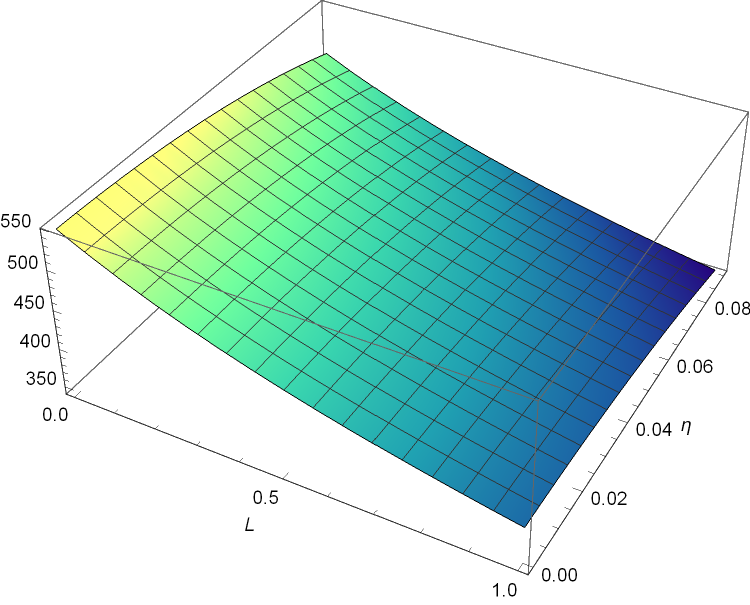}}\label{fig cosmohorizon}}
   \caption{Evolution of $r_h$ (left) and $r_c$ (right) w.r.t. $\eta$ and $L$. Here, $M=1$,  $\Lambda=10^{-5}$.}
   \label{fig horizon}
\end{figure}

Fig. \ref{fig horizon} shows the evolution of $r_h$ and $r_c$ with respect to the parameters $\eta$ and $L$. We observe that the event horizon increases monotonically with $\eta$, and it becomes progressively steeper, signifying a growing rate of increase. There is no substantial effect on $r_h$ by the LV parameter $L$, except for a slight increase in the radius with increasing $L$. From Fig. \ref{fig cosmohorizon}, we see that the effects of the parameters $\eta$ and $L$ on the cosmological horizon are much more significant than those of the event horizon. $r_c$ is reduced with increasing $L$ and $\eta$ though the effect is more prominent with the rise of $L$. The region enclosed by the event horizon and the cosmological horizon, where $f(r)>0$ is the domain of outer communication. The event horizon increases slightly with $\eta$ and $L$, and the cosmological horizon, showing a notable reduction with greater $\eta$ and $L$ indicates that the region of outer communication is reduced with the rise of the GM parameter and the LV parameter. We will consider the radial coordinate range, which is in the region of outer communication, since the motion of test particles around a black hole occurs in this region.

\section{Geodesic Motions} 
We examine the geodesic structure of lightlike and timelike particles in the spacetime of a Schwarzschild-dS-like BH with GM. We analyze the influence of the parameters describing the underlying spacetime geometry on features like particle trajectories, dynamics of test particles, photon sphere, shadow size, conditions for circular orbits, etc.,\\ 
The Lagrangian of geodesics is given by 
\begin{align}\label{eqn lagran}
2 \mathscr{L}= f(r) \dot{t}^2 -\dfrac{(1+L) \dot{r}^2}{f(r)}- r^2 \dot{\theta}^2 - r^2 \sin^2\theta \dot{\varphi}^2,
\end{align}
where the overdot denotes differentiation with respect to the proper time parameter $\tau$. The geodesic motion is investigated in the equatorial plane described by $\theta=\dfrac{\pi}{2}$ and $\dot{\theta}=0$, due to the spherical symmetry of the considered spacetime.
The components of momentum are given by
\begin{align}\label{eqn components}
p_r=-\dfrac{1+L}{f(r)}\dot{r},\hspace{0.4cm} p_t=f(r)\dot{t}\equiv E, \hspace{0.4cm} p_\phi=-r^2\dot{\phi}\equiv -\Lb.
\end{align}
where $E$ and $\Lb$ are the particle's conserved total energy and angular momentum, respectively.
Using Eq. \eqref{eqn components} in Eq. \eqref{eqn lagran}, the radial equation for time-like or null geodesic is given by
 \begin{align}\label{eqn radial}
\dot{r}^2+V_{eff}(r)= \mathscr{E},
\end{align}
where $\mathscr{E}=\dfrac{E^2}{1+L}$ is the effective energy of the particle. The effective potential $V_{eff}(r)$ for the radial equation is given by 
\begin{align}\label{eqn common veff}
V_{eff}(r)=\bigg(\delta+\dfrac{\Lb^2}{r^2}\bigg)\dfrac{f(r)}{1+L}=\dfrac{1}{1+L}\bigg(\delta+\dfrac{\Lb^2}{r^2}\bigg)\bigg(1-\kappa \eta^2-\dfrac{2M}{r}-\dfrac{(1+L)\Lambda r^2}{3}\bigg),
\end{align}
where $\delta=0,1$ for lightlike and timelike particles, respectively. We notice that various parameters affect the effective potential for both the timelike and null geodesics, including the symmetry breaking parameter $\eta$, LV parameter $L$, the cosmological constant $\Lambda$ and the angular momentum $\Lb$. These parameters collectively influence the gravitational field of the spacetime, thereby altering the dynamics of test particles. 
Next, we examine the motion of photons and timelike particles, respectively.

\section{Null geodesics}
The paths traversed by massless photon particles in curved spacetime are described by lightlike or null geodesics \cite{Ahmad2025}. The effective potential for null geodesics is given by
 \begin{align}\label{eqn nullveff}
V_{eff}(r)=\dfrac{1}{(1+L)}\dfrac{\Lb^2}{r^2}\bigg(1-\kappa \eta^2-\dfrac{2M}{r}-\dfrac{(1+L)\Lambda r^2}{3}\bigg).
\end{align} 

\begin{figure}[h!]
\centering
  \subfloat[\centering ]{{\includegraphics[width=170pt,height=160pt]{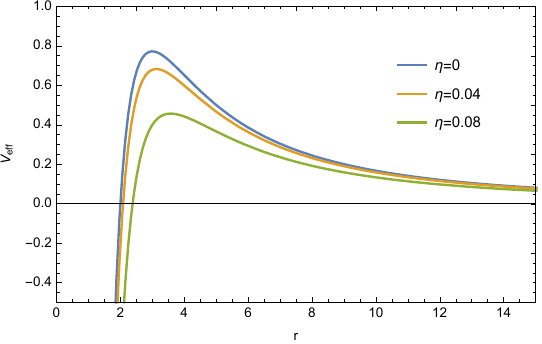}}\label{fig VetaNULL}}
  \qquad
   \subfloat[\centering ]{{\includegraphics[width=170pt,height=160pt]{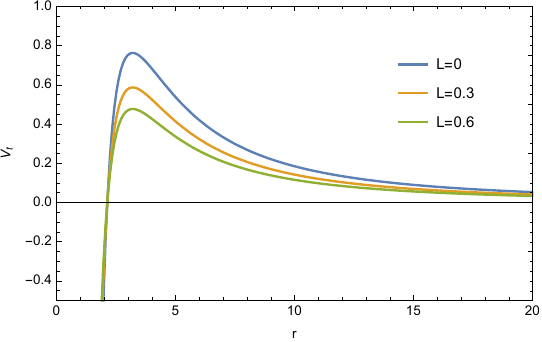}}\label{fig VLNULL}}
   \caption{Behaviour of $V_{eff}$ as a function of $L$ (right) and $\eta$ (left) with  $M=1$,  $\Lb=5$, $\Lambda=10^{-5}$. Here, (a) $L=0.2$ and (b) $\eta=0.05$.}
   \label{fig VNULL}
\end{figure}

In Fig. \ref{fig VNULL}, we illustrate the variation of the effective potential of null geodesics w.r.t. the radial coordinate $r$, for various values of the GM parameter $\eta$ and the LV parameter $L$. The effective potential is zero at the event horizon, rises to a maximum and then decays towards the cosmological horizon. We notice from Figs. \ref{fig VetaNULL} and \ref{fig VLNULL}, the maxima of the effective potential, which corresponds to the photon sphere radius, reduces with the rise of the parameters $L$ and $\eta$. This suggests that higher values of the GM parameter and stronger LV lower the potential barrier, which indicates that the GM and LV parameters contribute a repulsive nature in the spacetime. The peak of the potential at $r=r_{ph}$, also corresponds to the unstable circular orbit of photons. here, $r=r_{ph}$ represents the radius of the photon sphere \cite{Mohaddese 2022}. The peaks moving to the right with the increase of $\eta$ suggest the shifting away of the unstable circular orbits from the central object with the rise of the GM parameter.

\begin{figure}[h!]
\centering
  \subfloat[\centering ]{{\includegraphics[width=170pt,height=160pt]{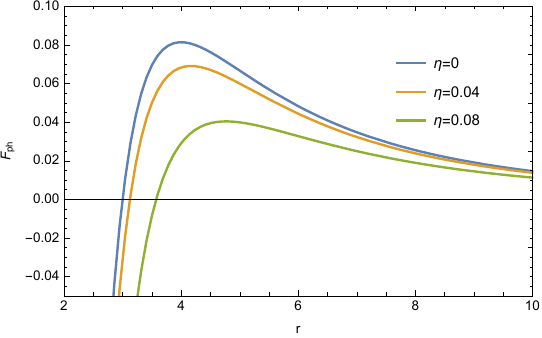}}\label{fig ForceEta}}
  \qquad
   \subfloat[\centering ]{{\includegraphics[width=170pt,height=160pt]{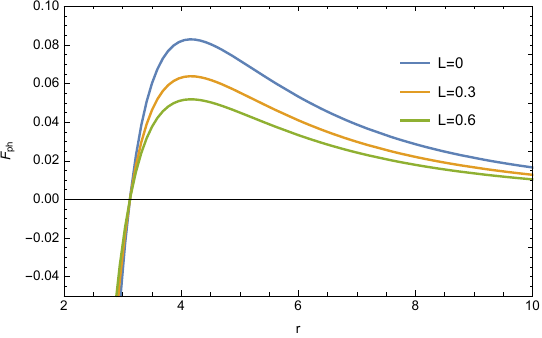}}\label{fig ForceL}}
   \caption{Force on photon particles as a function of  $r$ for different choices of $\eta$ (left) and $L$ (right) with  $M=1$, $\Lb=5$ and $\Lambda=10^{-5}$. Here, (a) $L=0.2$ and (b) $\eta=0.05$.}
   \label{fig ForcePhoton}
\end{figure}

Now, we determine the effective force on the massless photon particles in the gravitational field of the spacetime. This force is obtained from the effective potential gradient associated with null geodesics, and it sheds light on the nature of the
gravitational interaction in the spacetime \cite{Faizuddin2026}. The influence of the GM parameter and the LV parameter on the gravitational force is examined. The force on photons for null geodesics is given by 
\cite{Ahmad2025}
\begin{eqnarray}
F_{p}(r)= -\dfrac{1}{2}\dfrac{dV_{eff}(r)}{dr}.
\end{eqnarray}
Using the expression of the effective potential from Eq. \eqref{eqn nullveff}, we find
\begin{eqnarray}\label{eqn force photon}
F_{p}(r)=\dfrac{\Lb^2}{r^3(1+L)}\bigg[1- \kappa \eta^2-\dfrac{3M}{r}\bigg].
\end{eqnarray}
From Eq. \eqref{eqn force photon}, we see that several parameters influence the effective force on the lightlike particles, including the GM parameter $\eta$, the LV parameter $L$ and the angular momentum $\Lb$. Furthermore, in the limit $L=0$, corresponding to the absence of LV, the force reduces as 
\begin{eqnarray}\label{eqn ForceSdS}
F_{p}(r)=\dfrac{\Lb^2}{r^3}\bigg[1- \kappa \eta^2-\dfrac{3M}{r}\bigg],
\end{eqnarray}   
which is the force in the Schwarzschild-dS BH spacetime with GM. If $\eta=0$ Eq. \eqref{eqn ForceSdS} gives 
\begin{eqnarray}\label{eqn ForceSdS0}
F_{p}(r)=\dfrac{\Lb^2}{r^3}\bigg[1-\dfrac{3M}{r}\bigg],
\end{eqnarray} 
which corresponds to the force in the Schwarzschild BH spacetime. 
In Fig. \ref{fig ForcePhoton}, we illustrate the effective radial force on the photon particles in the gravitational field of the spacetime as a function of $r$, with varying $\eta$ and $L$. The effective force is zero at the photon sphere radius, $r_{ph}$, which depicts the position of the stationary point for circular geodesics. The force is negative for $r<r_{ph}$ and for $r>r_{ph}$, the force becomes positive. Here, the photons are deflected rather than pulled into the BH, as depicted in Figs. \ref{fig ForceEta} and \ref{fig ForceL}. The effective force exhibits a decreasing trend with larger $\eta$ and $L$. This indicates that stronger GM parameter and the LV parameter weaken the effective gravitational field of the spacetime, thereby reducing the effective force experienced by the photon particles. It ultimately leads to a less intense deflection of the photon particles.

Now, we study the motion of photons around the BH. The motion of particles is governed by Eq. \eqref{eqn radial}, and it is dependent on the effective potential and the impact parameter.  The photon sphere radius $r_{ph}$ can be obtained using the conditions $\dot{r}=0$ and $\ddot{r}=0$ \cite{MouXu2025, Roshila 2026}. Using these conditions, we get 
\begin{align}\label{eqn rr.}
V_{eff}(r)=\mathscr{E}, \hspace{0.6 cm} V_{eff}'(r)=0,
\end{align} 
which provides the radius of the photon sphere, $r=r_{ph}$ from the relation $ 2 f(r)=  f'(r)r$ and the critical impact parameter at $r=r_{ph}$ as
\begin{align}\label{eqn impactparameter}
b_{ph}=\dfrac{\Lb_{ph}}{E_{ph}}=\dfrac{r_{ph}}{\sqrt{f(r_{ph})}}=\dfrac{3M}{(1-\kappa \eta^2)\sqrt{1- \kappa \eta^2-\dfrac{2}{3}(1-\kappa \eta^2)-\dfrac{3 (1+L) M^2\Lambda}{(1-\kappa \eta^2)^2}}},
\end{align} where $r_{ph}={3M}/{(1-\kappa \eta^2)}$.
We can classify different kinds of motions of photons according to the value of the impact parameter $b$, assuming the movement of light rays in a radially inward direction. When $b= b_{ph}$, the photon particles start falling inward, and at $r=r_{ph}$, the photon particles spiral around the BH in circular orbits. The trajectories with $b<b_{ph}$ approach the BH and then falls into it since the potential barrier is not encountered. The trajectories do not encounter a turning point. For trajectories with $b>b_{ph}$, photons starting from $r>r_{ph}$ are pushed back by the potential barrier and escape to infinity, but photons with $r<r_{ph}$ plunge into the singularity. Table \ref{tab null} presents the numerical results of the radii of the event horizon, cosmological horizon, photon sphere and the impact parameter for different values of $\eta$ and $L$. At a fixed value of the LV parameter $L$, $r_h$ , the photon sphere radius and the corresponding impact parameter increase with the rise of $\eta$ but the radius of the cosmological horizon reduces significantly. There is a reduction in the radial distance enclosed by the event horizon and the cosmological horizon, which is the outer communication region. For increasing $L$ with fixed $\eta$, the event horizon radius and the impact parameter increase slightly, but the cosmological horizon shows a notable reduction. The photon sphere radius is not affected by $L$. This means that with the rise of $L$, the outer communication region also becomes narrower.

\begin{table}[h]
 \centering
    \begin{tabular}{c c c c c c}
    \toprule
    $L$   &  $\eta$   &   $r_h $ &  $r_c$ &  $r_{ph}$ &$b_{ph}$ \\  
\hline
0.2    & 0 & 2.00003  &498.997 & 3 & 5.19643  \\
0.2 & 0.04 & 2.08383 &  488.799 & 3.12569  & 5.52644  \\
0.2    & 0.08 & 2.38343  &456.829 & 3.57504 & 6.76024  \\
\hline
0    & 0.05 & 2.13412  &529.166 & 3.20113 & 5.72769  \\
0.3 & 0.05 & 2.13413 &  463.977 & 3.20113  & 5.72778  \\
0.6    & 0.05 & 2.13414  &418.117 & 3.20113 & 5.72788  \\
\hline
 \end{tabular}
\caption{$r_h$, $r_c$, $r_{ph}$ and $b_{ph}$ of the photon sphere with varying $\eta$ and $L$. }
    \label{tab null}
\end{table}

Now, we investigate the photon trajectories in the proximity of the BH. We focus on the geometry of null geodesics under the influence of the background gravitational field. To study the path of photons, the relation between $\phi$ and $r$ is required, which is obtained from Eqs. \eqref{eqn components} and \eqref{eqn radial} as 
\begin{eqnarray}\label{eqn pho}
\dfrac{\dot{r}^2}{\dot{\phi}^2}=\bigg(\dfrac{dr}{d\phi}\bigg)^2=r^4\bigg(\dfrac{1}{b^2(1+L)}-\dfrac{1}{r^2(1+L)}f(r)\bigg).
\end{eqnarray} 
 Using the transformation $u=\dfrac{1}{r}$, Eq. \eqref{eqn pho} can we written as 
\begin{eqnarray}\label{eqn photon}
\bigg(\dfrac{du}{d\phi}\bigg)^2=\dfrac{1}{b^2(1+L)}-\dfrac{u^2}{1+L}\bigg(1-\kappa \eta^2 -2 M u-\dfrac{(1+L)\Lambda}{3 u^2}\bigg) \equiv \Phi(u) .
\end{eqnarray} The trajectory of photons in the gravitational field of the spacetime is described by Eq. \eqref{eqn photon}. 

\begin{figure}[h!]
\centering
  \subfloat[\centering ]{{\includegraphics[width=150pt,height=150pt]{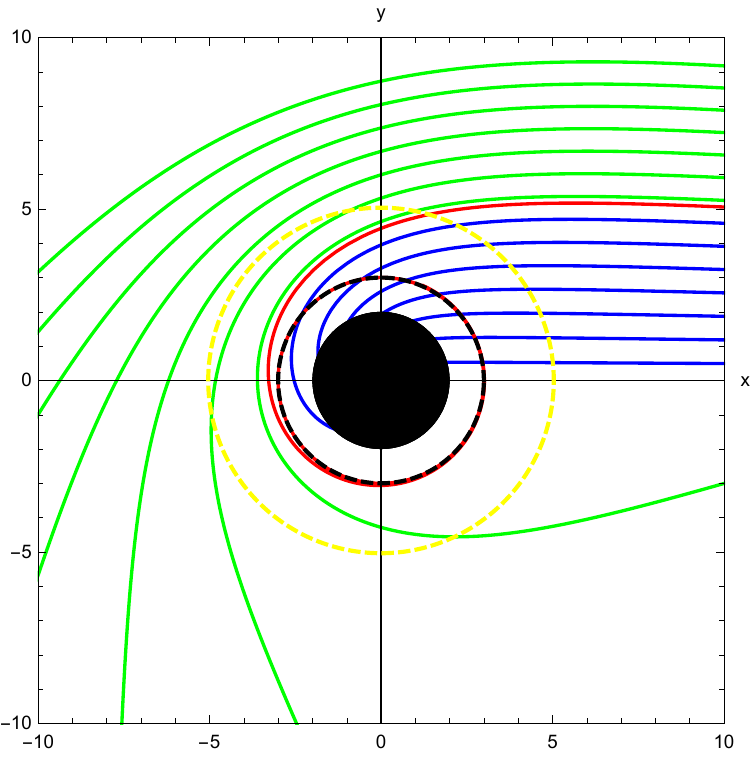}}\label{fig 1PhotraEta0}}
  \quad
   \subfloat[\centering ]{{\includegraphics[width=150pt,height=150pt]{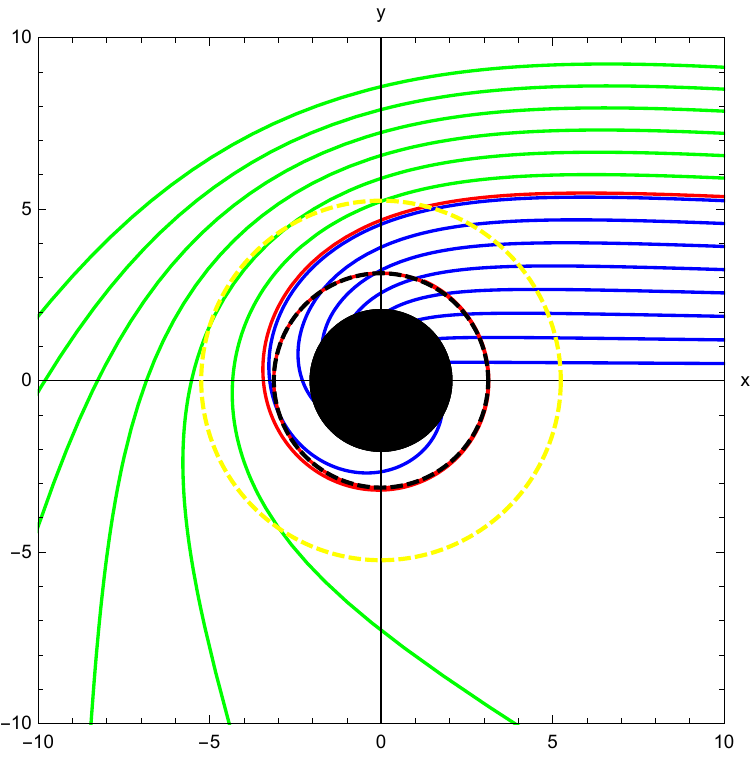}}\label{fig 1PhotraEta04}}
   \quad
   \subfloat[\centering ]{{\includegraphics[width=150pt,height=150pt]{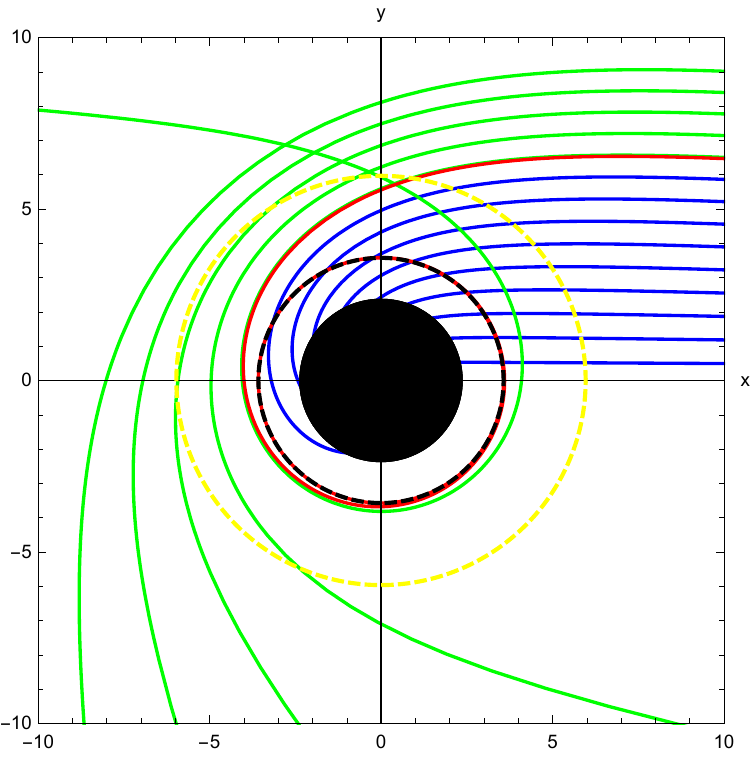}}\label{fig 1PhotraEta08}}
   \caption{Photon trajectories around the BH varying $\eta$, with $M=1$, $L=0.2$, $\Lb=5$ and $\Lambda=10^{-5}$. The green, red and blue curves correspond to trajectories with $b > b_{ph}$, $b = b_{ph}$ and $b < b_{ph}$.}
   \label{fig PhotontraEt}
\end{figure}

\begin{figure}[h!]
\centering
  \subfloat[\centering ]{{\includegraphics[width=150pt,height=150pt]{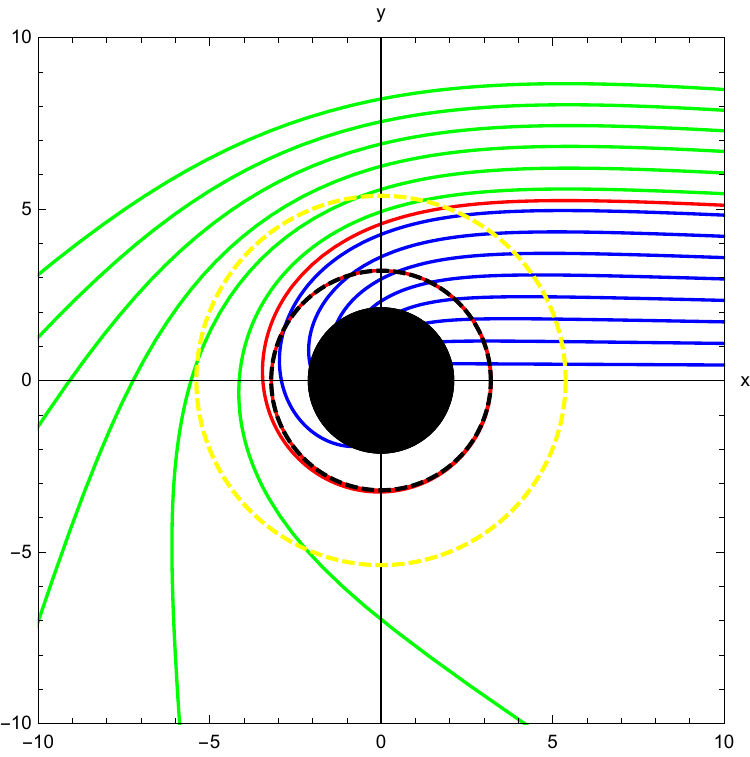}}\label{fig PhotraL0}}
  \quad
   \subfloat[\centering ]{{\includegraphics[width=150pt,height=150pt]{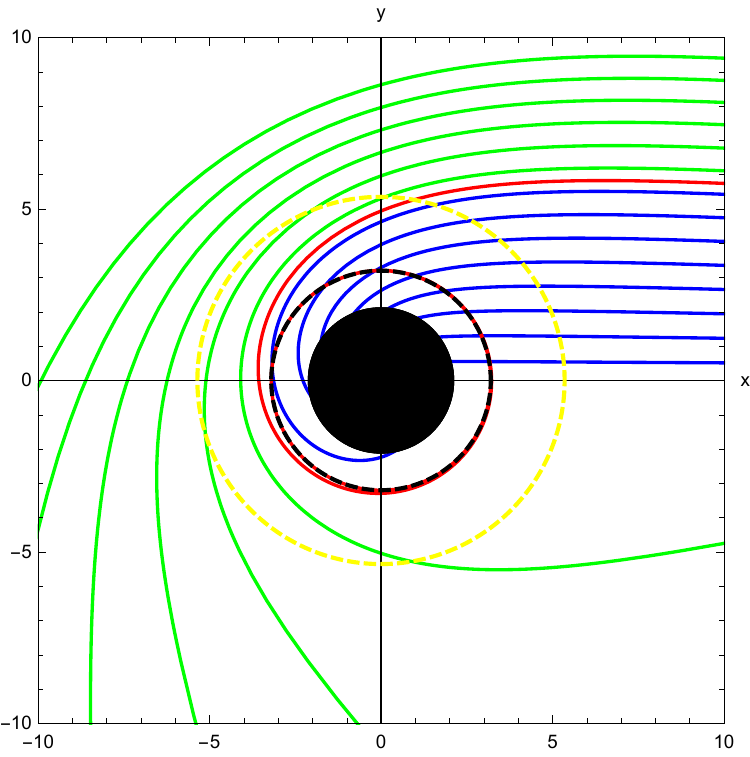}}\label{fig PhotraL03}}
   \quad
   \subfloat[\centering ]{{\includegraphics[width=150pt,height=150pt]{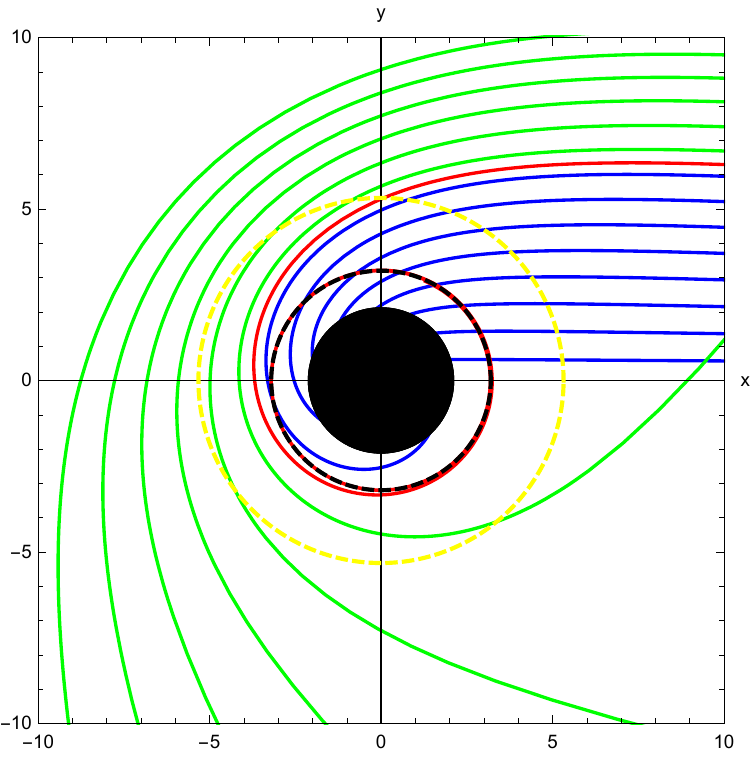}}\label{fig PhotraL06}}
   \caption{Photon trajectories around the BH varying $L$, with $M=1$, $\Lb=5$, $\eta=0.05$ and $\Lambda=10^{-5}$. The green, red and blue curves correspond to trajectories with $b > b_{ph}$, $b = b_{ph}$ and $b < b_{ph}$.}
   \label{fig PhotraL}
\end{figure}

When $\eta=0$, $\Lambda=0$ and $L=0$, Eq. \eqref{eqn photon} reduces to $({du}/{d\phi} )^2={1}/{b^2}-u^3+2 M u^3$, which is the equation of photon motion in the Schwarzschild BH spacetime. Using Eq. \eqref{eqn photon}, we numerically plot the photon trajectories of the spacetime in Figs. \ref{fig PhotontraEt} and \ref{fig PhotraL}. The observer is placed in the region enclosed by the cosmological horizon and the event horizon to examine the photon trajectory. In the figures, the blue light rays with $b<b_{ph}$ plunge into the BH. The red photon ray with $b=b_{ph}$ has an unstable circular orbit at $r=r_{ph}$ and asymptotically circles the BH at the photon sphere. The green light rays corresponding to $b>b_{ph}$ encounter the potential barrier and are deflected at the turning point $u_0$. $u_0={1}/{r_0}$ satisfies $\Phi(u_0)=0$. The radial position $u_0$ is essential for establishing the photon trajectories; it determines the geodesic structure. The event horizon is illustrated as black disks in the plots, the black dashed circle denotes the photon sphere, and the yellow dashed circle depicts the corresponding shadows of the black hole. From Fig. \ref{fig PhotontraEt}, we observe a substantial impact on the photon path following an increase in the GM parameter $\eta$. We observe a notable contraction on the path of light with the rise of $\eta$. Also, more rays are captured by the BH due to the increased impact parameter, resulting from higher GM parameter values. This results in a larger shadow size as seen by distant observers as depicted by the yellow dashed circle. Additionally, we observe that the event horizon expands with higher $\eta$ values, which aligns with the numerical values given in Table \ref{tab null}. For Fig. \ref{fig PhotraL}, we do not observe a significant effect with the rise of the $L$ since the radius of the photon sphere is not affected by the LV parameter, and there is only a slight increase in the impact factor with the rise of $L$.

\subsection{Lyapunov Stability} 
\begin{figure}[h!]
\centering
  \subfloat[\centering ]{{\includegraphics[width=170pt,height=160pt]{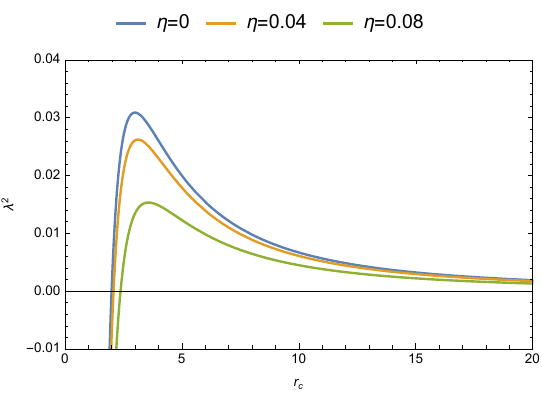}}\label{fig NullLyaeta}}
  \qquad
   \subfloat[\centering ]{{\includegraphics[width=170pt,height=160pt]{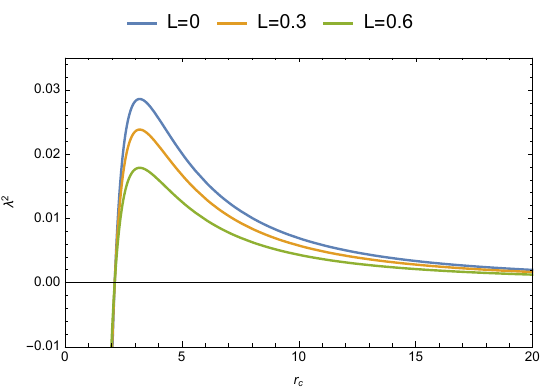}}\label{fig NullLyaL}}
   \caption{Dependence of $\lambda^2$ on $\eta$ (a) and $L$ (b) with $M=1$, $\Lb=5$, $\Lambda=10^{-5}$. Here, (a) $L=0.2$ and (b) $\eta=0.05$.}
   \label{fig NullLya}
\end{figure}

We will analyze the stability of circular orbits for null geodesics using the Lyapunov exponent. The Lyapunov exponent represents the average divergence rate of neighbouring geodesics in phase space \cite{Mrinnoy M 2024}. Negative values of the Lyapunov exponent suggest the convergence of the neighbouring orbits, while positive values imply the divergence of neighbouring geodesics, which results in chaos in the system \cite{Ruifang2024}.  The Lyapunov exponent is expressed as \cite{Ahmad2025, Roshila 2026}
\begin{eqnarray}\label{eqn Lya}
\lambda=\sqrt{-\dfrac{V_{eff}''(r)}{2 \dot{t}^2}}\bigg\vert_{r=r_{ph}},
\end{eqnarray} 
where $\dot{t}=\dfrac{E}{f(r)}$. Using Eq. \eqref{eqn nullveff}  in the above equation and simplifying , we have
\begin{eqnarray}\label{eqn Lyanull}
\lambda^2=\dfrac{(-1+8 \pi \eta^2)(6M +r(-3+24 \pi \eta^2+r^2 \Lambda+L r^2 \Lambda))}{3(1+L)r^3}\bigg\vert_{r=r_{ph}}.
\end{eqnarray}
From Eq. \eqref{eqn Lyanull}, we see that the Lyapunov exponent is affected by several factors, including the cosmological constant $\Lambda$, the LV parameter $L$ and the GM parameter $\eta$.
The circular geodesics are unstable for $\lambda>0$, stable for $\lambda<0$ and marginally stable when $\lambda=0$. Fig. \ref{fig NullLya} presents the dependence of $\lambda^2$ on $r_{ph}$ for null circular orbits. We notice that $\lambda^2$ retains positive, real values for suitable choices of $\eta$ and $L$, suggesting the unstable behaviour of the null circular orbits. A larger $\lambda$ corresponds to higher instability. We also notice that the instability reduces with higher values of $\eta$ and $L$. The GM parameter introduces a repulsive contribution, which reduces instability by going against the gravitational attraction.

\section{Timelike geodesics}
\begin{figure}[h!]
\centering
  \subfloat[\centering ]{{\includegraphics[width=170pt,height=160pt]{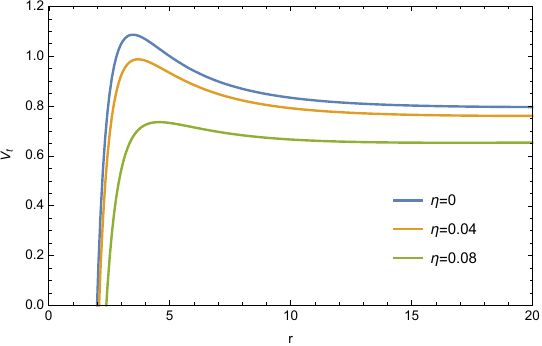}}\label{fig TimeVeffeta}}
  \qquad
   \subfloat[\centering ]{{\includegraphics[width=170pt,height=160pt]{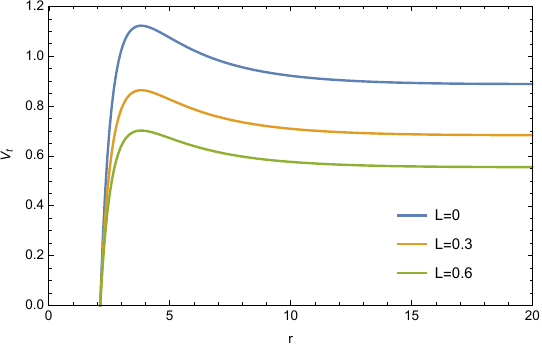}}\label{fig TimeVeffL}}
   \caption{Behaviour of $V_t$ w.r.t. $\eta$ (a) and $L$ (b) with $M=1$, $\Lb=5$, $\Lambda=10^{-5}$. Here,(a) $L=0.2$ and (b) $\eta=0.05$.}
   \label{fig TimeVeff}
\end{figure}

\begin{figure}[h!]
\centering
  \subfloat[\centering ]{{\includegraphics[width=170pt,height=160pt]
   {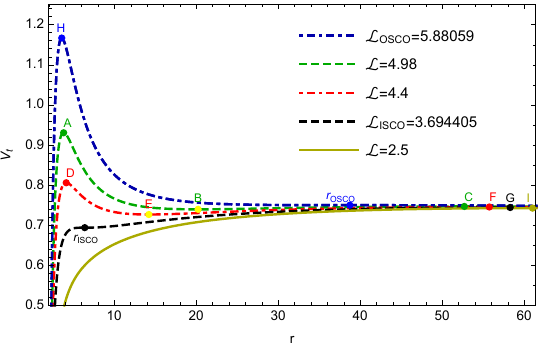}}\label{fig VtimeL1big}}
  \qquad
   \subfloat[\centering ]{{\includegraphics[width=170pt,height=160pt]{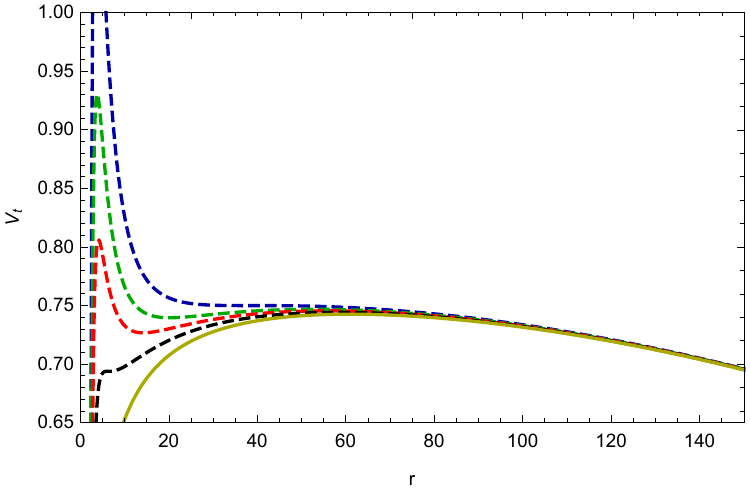}}\label{fig VtimeL1}}
   \caption{Behaviour of $V_t$ as $\mathcal{L}$ varies, with  $M=1$, $\Lambda=10^{-5}$, $\eta=0.05$, $L=0.2$ (left) and the effective potential in the far and near regions from the BH for varying $\Lb$ (right).}
   \label{fig TimeVeffL}
\end{figure}

Here, we investigate the time-like geodesics in the Schwarzschild-dS-like BH spacetime with GM. The radial motion and timelike orbit stability are governed by the effective potential. The influence of the curvature of the spacetime on the particle motion is also encapsulated by the effective potential. The effective potential for timelike geodesics is given by
\begin{eqnarray}\label{eqn Veff time}
V_{eff}(r)=\dfrac{1}{(1+L)}\bigg(1+\dfrac{\Lb^2}{r^2}\bigg)\bigg(1-\kappa \eta^2-\dfrac{2M}{r}-\dfrac{(1+L)\Lambda r^2}{3}\bigg).
\end{eqnarray}
The radial dependence of the effective potential for various values of the GM parameter $\eta$ and LV parameter $L$ are illustrated in Fig. \ref{fig TimeVeff}. Circular orbits are associated with the effective potential extrema. The minimum and maximum exhibited by the effective potential curve characterize the stable and unstable circular orbits, while the ISCO is marked by the inflection point where ${d^2V_{eff}(r)}/{dr^2}=0$ \cite{Shokhzod2025}.  We observe that the maxima of the potential is lowered with larger $\eta$ and $L$. This indicates that a particle requires less energy to be in an unstable circular orbit around the BH with the rise of the GM parameter and the LV parameter. Furthermore, we notice from Fig.\ref{fig TimeVeffeta} that the potential peak shifts farther to the right with the rise of $\eta$, suggesting that the unstable circular orbits move farther away from the BH.\\
The different possible orbits for timelike geodesics can be characterized by using the angular momentum $\Lb$. Different orbit types can be obtained by using different angular momenta. To elucidate, we also consider the points which simultaneously satisfy the conditions $V_{eff}'(r)=0$ and $V_{eff}''(r)=0$. The associated angular momenta are then classified as $\Lb_{ISCO}$ and $\Lb_{OSCO}$ where the ISCO and the OSCO occur. \\
In Fig. \ref{fig VtimeL1big}, we illustrate the effective potential for various choices of the angular momentum $\Lb$. From the figure, we observe that for $\Lb$ less than $\Lb_{ISCO}$, the effective potential has a maximum at a larger radial distance. It is the point $I$ in the figure, and it corresponds to the only unstable circular orbit for $\Lb=2.5$. When $\Lb=\Lb_{ISCO}$, the effective potential has an unstable circular orbit and the ISCO, denoted by points $G$ and $r_{ISCO}$ on the plot. When $\Lb_{ISCO}<\Lb<\Lb_{OSCO}$, the effective potential exhibits two maxima and one minimum, which are associated with two unstable circular orbits and a stable circular orbit. For $\Lb=4.4$, the two maxima are denoted by points $D$ and $F$, and the stable orbit by point $E$. The stable circular orbits are denoted by the yellow dots. We also notice that the stable circular orbit moves away from the BH as the value of angular momentum increases, and the unstable circular orbit moves towards the BH with the rise of the angular momentum value. For $\Lb>L_{OSCO}$, the effective potential features an unstable circular orbit and the OSCO, denoted by points $H$ and $r_{OSCO}$ respectively. For $\Lb>\Lb_{OSCO}$, the particle would feature an unstable circular orbit corresponding to a single peak. Furthermore, the peak of the curves goes higher with the rise of the angular momentum $\Lb$, meaning higher energy is needed by the particles to sustain their motion.

\begin{figure}[h!]
\centering
  \subfloat[\centering ]{{\includegraphics[width=150pt,height=150pt]{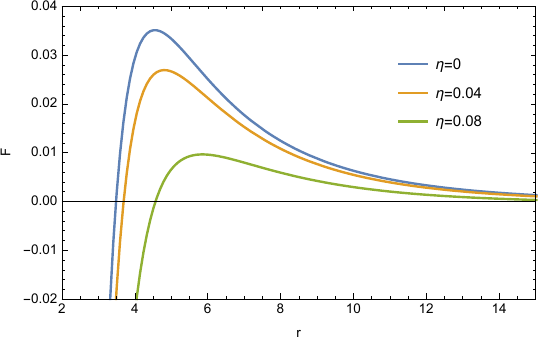}}\label{fig ForcetimeEta}}
  \quad
   \subfloat[\centering ]{{\includegraphics[width=150pt,height=150pt]{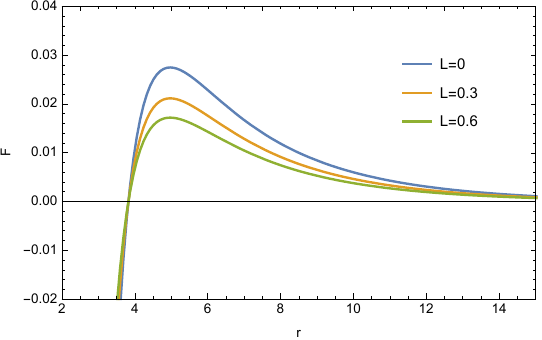}}\label{fig ForcetimeL}}
   \quad
   \subfloat[\centering ]{{\includegraphics[width=150pt,height=150pt]{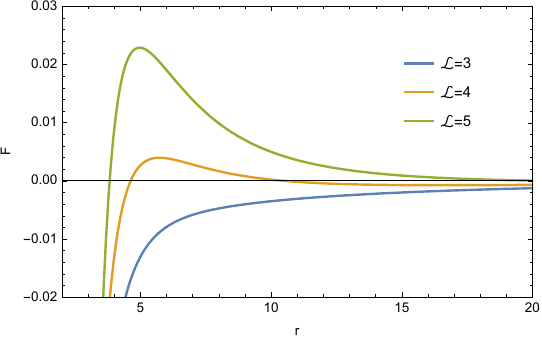}}\label{fig ForcetimeL1}}
   \caption{Illustration of timelike particle trajectories varying $L$ with  $\Lb=5$, $\Lambda=10^{-10}$, $\eta=0.05$ and $M=1$. Here, $L=0, 0.3$ and $0.6$ in panels (a), (b) and (c).}
   \label{fig ForceTime}
\end{figure}

Now, we discuss the effective force on the timelike particle. The effective force acting on the particle offers valuable insight into the nature of the particle's motion, whether it experiences attraction to or repulsion from the BH. The effective force on timelike particles is defined as \cite{G. Mustafa 2026}
\begin{eqnarray}
F(r)&= &-\dfrac{1}{2}\dfrac{dV_{eff}(r)}{dr} \nonumber\\
&=&\dfrac{1}{1+L}\bigg(-\dfrac{M}{r^2}+\dfrac{(1+L) \Lambda r}{3}+\dfrac{\Lb^2}{r^3}-\dfrac{\Lb^2 \kappa \eta^2}{r^3}-\dfrac{3 M \Lb^2}{r^4}\bigg).
\end{eqnarray} 
In the above expression, the second and third terms provide a repulsive contribution to the effective force, while the remaining terms represent attractive contributions. In Fig. \ref{fig ForceTime}, we display a set of plots depicting the radial dependence of the effective force for different choices of $\eta$, $L$ and $\Lb$. The influence of increasing the GM and LV parameters is reflected in the weakening of the effective force acting on 
the massive particles. The inclusion of the GM and the LV parameters causes the effective force to become less repulsive. Meanwhile, increasing the specific angular momentum $\Lb$ enhances the total force on the massive particles.

\subsection{Circular orbits}
\begin{figure}[h!]
\centering
  \subfloat[\centering ]{{\includegraphics[width=170pt,height=160pt]{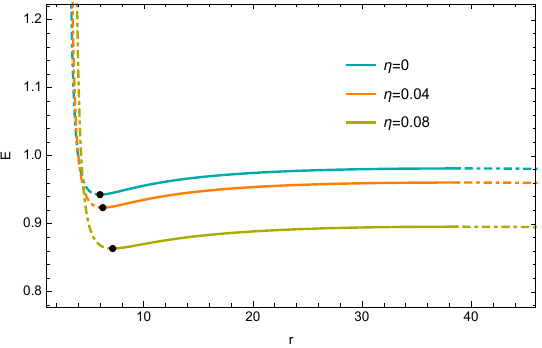}}\label{fig SEnergyeta}}
  \qquad
   \subfloat[\centering ]{{\includegraphics[width=170pt,height=160pt]{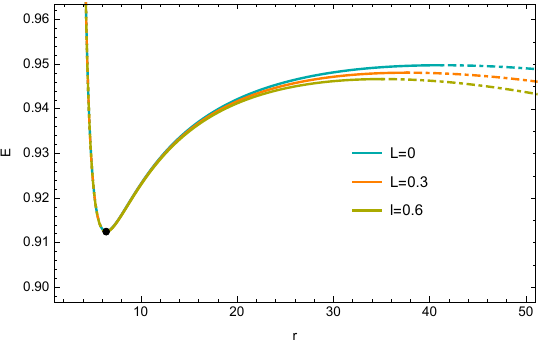}}\label{fig SEnergyL}}
   \caption{Behaviour of $E$ as a function of $L$ (right) and $\eta$ (left) with $M=1$, $\Lb=5$, $\Lambda=10^{-5}$. Here,(a) $L=0.2$ and (b) $\eta=0.05$.  The solid and dotdashed portions of the curve correspond to the positions of stable and unstable circular orbits, respectively.}
   \label{fig SEnergyE}
\end{figure}

\begin{figure}[h!]
\centering
  \subfloat[\centering ]{{\includegraphics[width=170pt,height=160pt]{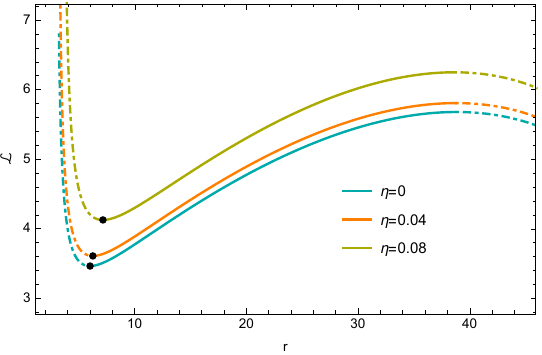}}\label{fig SAngulareta}}
  \qquad
   \subfloat[\centering ]{{\includegraphics[width=170pt,height=160pt]{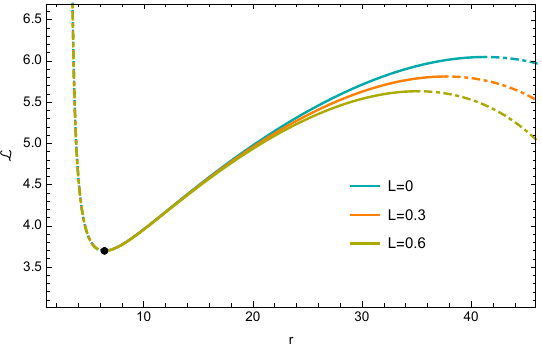}}\label{fig SAngularL}}
   \caption{Behaviour of $\Lb$ as a function of  $\eta$ (a) and $L$ (b) with $M=1$, $\Lb=5$, $\Lambda=10^{-5}$. Here,(a) $L=0.2$ and (b) $\eta=0.05$. The solid and dotdashed portions of the curve correspond to the positions of stable and unstable circular orbits, respectively.}
   \label{fig SAngularM}
\end{figure}

To study the dynamics of massive particles in the background spacetime, we will discuss the circular orbits, ISCO and OSCO. The two conditions that should be satisfied by particles to maintain a circular orbit are given by \cite{	Shokhzod2025}
\begin{eqnarray}
V_{eff}(r)=\mathscr{E}\equiv \dfrac{E^2}{1+L}, \hspace{0.6 cm} \dfrac{d V_{eff(r)}}{dr}=0.
\end{eqnarray}
From the above two conditions, we obtain
\begin{align}\label{eqn L2}
\Lb^2&=\dfrac{\Lambda(1+L)r^5-3 M r^2}{-3 r+3 r \kappa~ \eta^2+9M},\\
E^2&=-\dfrac{(6 M+r(-3+3 \kappa ~\eta^2+r^2\Lambda+L r^2\Lambda))^2}{9 r(3M+r(-1+\kappa~ \eta^2))}, 
\end{align}where $E$ and $\Lb$ are the particle's total energy and specific angular momentum, associated with circular motion. 
Figs. \ref{fig SEnergyE} and \ref{fig SAngularM} illustrates the radial dependence of the energy $E$ and the angular momentum $\Lb$ for different values of $\eta$ and $L$. We observe from the figures that increasing $L$ reduces both the energy and angular momentum. This suggests that increasing the LV parameter weakens the gravitational potential and lowers the energy and angular momentum required by the particles to be in circular orbits. So, the motion of the particles is slower and orbits the black hole loosely. In contrast, we notice from Figs. \ref{fig SEnergyeta} and \ref{fig SAngulareta} that higher values of $\eta$ reduces the specific energy $E$ but enhances the specific angular momentum $\Lb$. This indicates that the presence of a stronger GM makes the circular orbits more energetically accessible, while simultaneously enhancing the angular momentum required by the orbits to remain stable.  \\
The solid and dotdashed portions of the curve are associated with the portions of stable and unstable circular orbits, respectively. The minimum and maximum of the $\Lb$ and $E$ curves are associated with the ISCO and the OSCO radii, respectively. Also, the ISCO radii are denoted by back dots in the figures. From the figures, we notice that increasing $\eta$ pushes the ISCO radius outward while the OSCO radius decreases slightly. Therefore, the stable orbit existing region becomes narrower with higher values of the GM parameter. Furthermore, increasing $L$ reduces the OSCO radii significantly, but makes no notable change in the ISCO radii, as seen from the graph. So, a higher LV parameter leads to the shrinking of the stable orbit existing region as well. 

\begin{figure}[h!]
\centering
  \subfloat[\centering ]{{\includegraphics[width=150pt,height=150pt]{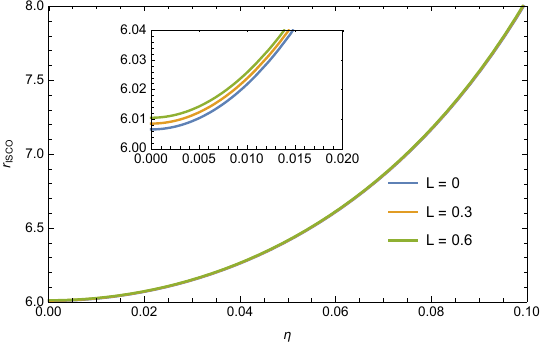}}\label{fig EtarISCOL}}
  \quad
   \subfloat[\centering ]{{\includegraphics[width=150pt,height=150pt]{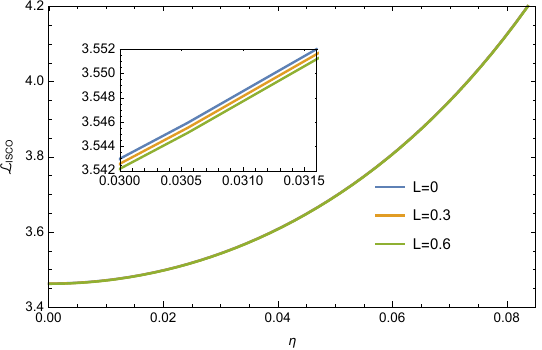}}\label{fig EtaLISCOL}}
   \quad
   \subfloat[\centering ]{{\includegraphics[width=150pt,height=150pt]{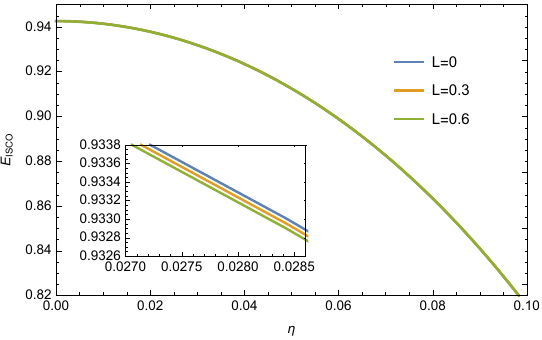}}\label{fig EtaEISCOL}}
   \centering
   \quad
  \subfloat[\centering ]{{\includegraphics[width=150pt,height=150pt]{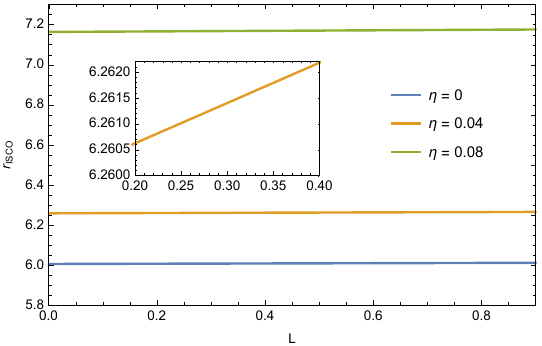}}\label{fig LrISCOEta}}
  \quad
   \subfloat[\centering ]{{\includegraphics[width=150pt,height=150pt]{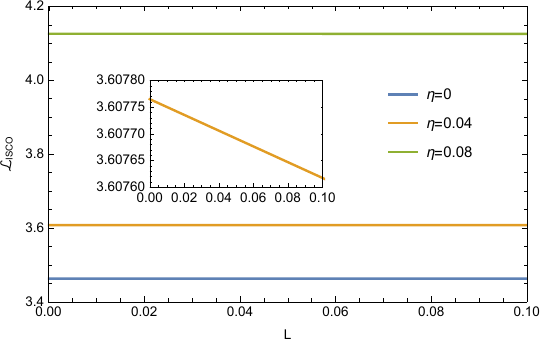}}\label{fig LLISCOeta}}
   \quad
   \subfloat[\centering ]{{\includegraphics[width=150pt,height=150pt]{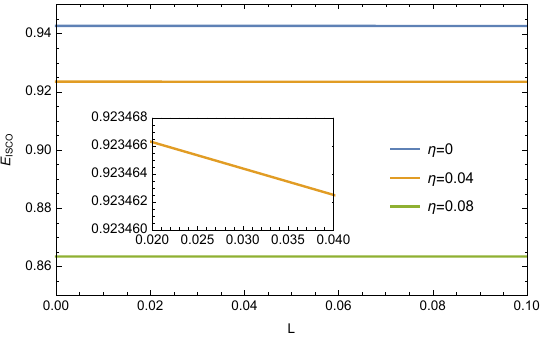}}\label{fig LEISCOEta}}
   \caption{Variation of $r_{ISCO}$, $L_{ISCO}$ and $E_{ISCO}$ w.r.t. $\eta$ for various choices of $L$. Here, $M=1$, $\Lb=5$, $\Lambda=10^{-5}$.}
   \label{fig ISCO}
\end{figure}

\begin{figure}[h!]
\centering
  \subfloat[\centering ]{{\includegraphics[width=150pt,height=150pt]{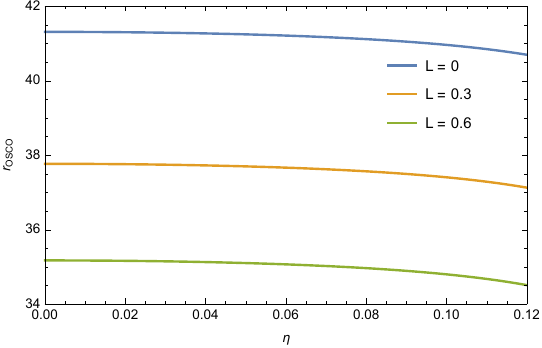}}\label{fig EtarOSCOL}}
  \quad
   \subfloat[\centering ]{{\includegraphics[width=150pt,height=150pt]{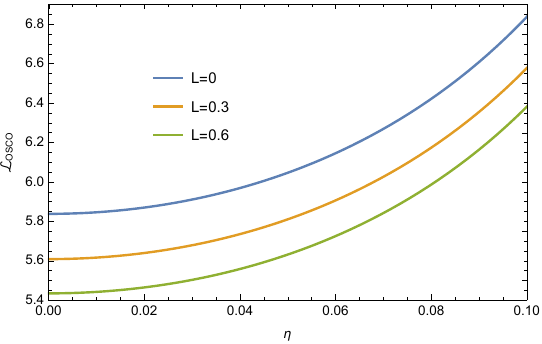}}\label{fig EtaLOSCOL}}
   \quad
   \subfloat[\centering ]{{\includegraphics[width=150pt,height=150pt]{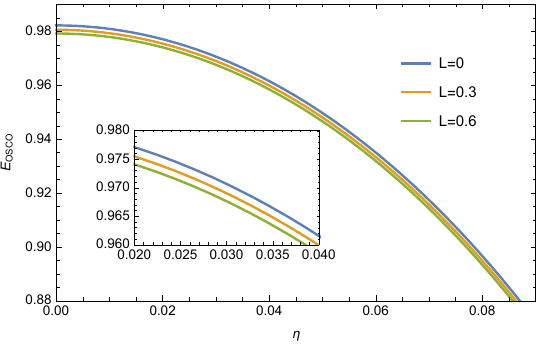}}\label{fig EtaEOSCOL}}

   \centering
  \subfloat[\centering ]{{\includegraphics[width=150pt,height=150pt]{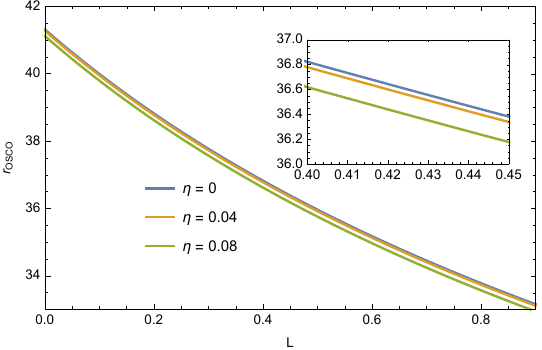}}\label{fig LrOSCOEta}}
  \quad
   \subfloat[\centering ]{{\includegraphics[width=150pt,height=150pt]{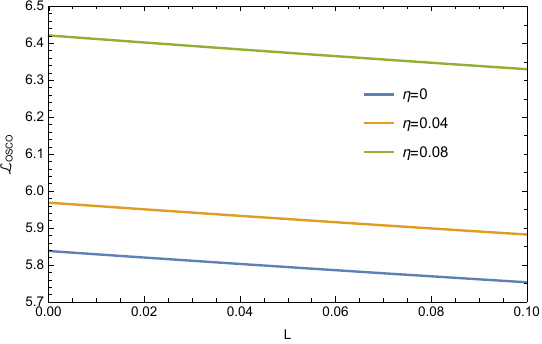}}\label{fig LLOSCOEta}}
   \quad
   \subfloat[\centering ]{{\includegraphics[width=150pt,height=150pt]{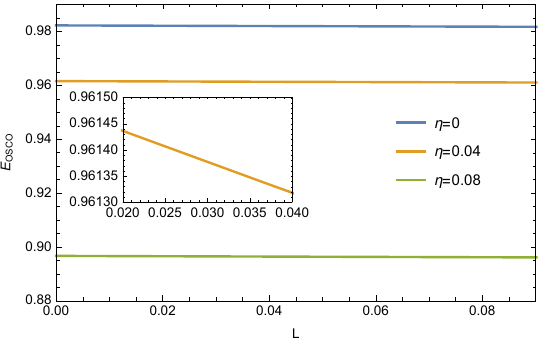}}\label{fig LEOSCOEta}}
   \caption{Variation of $r_{OSCO}$, $L_{OSCO}$ and $E_{OSCO}$ w.r.t.  $L$ for various values of $\eta$. Here, $M=1$, $\Lb=5$, $\Lambda=10^{-5}$.}
   \label{fig OSCO}
\end{figure}

The smallest stable circular orbit is marked by the ISCO. Particles in circular motion become dynamically unstable beyond the ISCO. For an orbit to be stable, the second derivative of the effective potential should satisfy the condition \cite{Shokhzod2025}
\begin{eqnarray}
V_{eff}'(r)\geq 0,
\end{eqnarray} where ISCO corresponds to the equality.
The conditions for ISCO/OSCO are \cite{Roshila 2026}
\begin{eqnarray}
V_{eff}(r)=\mathscr{E}, \hspace{0.4 cm} V_{eff}'(r)=0, \hspace{0.4 cm} V_{eff}''(r)=0.
\end{eqnarray}
Using the expression of the effective potential on the above conditions we get \cite{Roshila 2026},
\begin{eqnarray}
3 f'(r) f(r) +  f(r) f''(r) r - 2 f'(r)^2 r\vert_{r=r_{SCO}}=0.
\end{eqnarray}
Substituting the expression of $f(r)$ into the above equation we get the expression of ISCO/OSCO as
\begin{eqnarray}
2 r M(1-\kappa \eta^2)+10 M(1+L)\Lambda r^3-\dfrac{8}{3}(1-\kappa \eta^2)(1+L)\Lambda r^4-12 M^2=0.
\end{eqnarray}
The above equation has two imaginary roots and two positive real roots; the smaller real root corresponds to the ISCO, and the larger real positive root corresponds to the OSCO. Numerical values of quantities associated with ISCO and OSCO are presented in Table \ref{tab SCO} to explore the consequences of incorporating the GM and LV parameters on the stable circular orbit boundaries. Increasing $\eta$ increases the ISCO radius while reducing the OSCO radius, thereby narrowing the stable orbit existing region. Increasing $L$ also displays a similar effect on the ISCO and OSCO radii, displaying a shrinking of the region of stable circular orbits.  \\ 
Graphical analysis of the behaviour of the ISCO and OSCO radii, as well as their corresponding energy and angular momentum at ISCO and OSCO for varying $L$ and $\eta$ are presented in Figs. \ref{fig ISCO} and \ref{fig OSCO}. The upper panel of Fig. \ref{fig ISCO} illustrates the variation of $r_{ISCO}$, $\Lb_{ISCO}$ and $E_{ISCO}$ with $\eta$ for different choices of $L$. The lower panel shows variation of $r_{ISCO}$, $\Lb_{ISCO}$ and $E_{ISCO}$ with $L$ for different choices of $\eta$. A growth in both $r_{ISCO}$ and $\Lb_{ISCO}$ is observed towards higher $\eta$ regimes while $E_{ISCO}$ decreases with $\eta$. This behaviour highlights that the presence of the GM parameter makes massive particles orbit the black hole in stable circular orbits with a larger radius, and a larger angular momentum is needed for the particles to remain in these stable orbits. Furthermore, we observe that the ISCO radii decrease with higher values of $L$ while $\Lb_{ISCO}$ and $E_{ISCO}$ are reduced for larger $L$. This implies that the LV parameter influences the spacetime in a way that pushes the stable circular orbits farther from the black hole.\\
The upper panel of Fig. \ref{fig OSCO} presents the variation of the OSCO radius and the associated angular momentum and energy with $\eta$ for different choices of $L$. The lower presents the variation with respect to $L$ for different choices of $\eta$. It is observed that that the OSCO radius and $E_{OSCO}$ decrease with the rise of $\eta$ while $L_{OSCO}$ is enhanced with increasing $\eta$. Furthermore, $r_{OSCO}$ , $L_{OSCO}$ and $E_{OSCO}$ are reduced with increasing $L$. All the above observations align with the numerical values computed in Table \ref{tab SCO}.

\begin{table}[h]
 \centering
    \begin{tabular}{c c c c c c c c}
    \toprule
    $L$   &  $\eta$   &   $r_{ISCO} $ &  $r_{OSCO}$ &  $\Lb_{ISCO}$ &$\Lb_{OSCO}$ & $\mathscr{E}_{ISCO}$ &$\mathscr{E}_{OSCO}$\\  
\hline
0.2    & 0 & 6.00783  & 38.8174 & 3.4626 & 5.67682 & 0.860477 & 0.89564  \\
0.2 & 0.04 &6.26062 &  38.7749 & 3.60747  & 5.80473 &0.842975 &0.876743 \\
0.2    & 0.08 & 7.16597  &38.6179 & 4.12508 & 6.24755& 0.788123 & 0.817438  \\
\hline
0    & 0.05 &6.41073  &41.2453 & 3.69473 & 6.04601& 0.91251 & 0.949727  \\
0.3 & 0.05 & 6.41329 & 37.6999 & 3.69424  &5.80974 & 0.800271 & 0.831499  \\
0.6    & 0.05 & 6.41587  &35.1035 & 3.69375 & 5.6309 & 0.721307 & 0.748374  \\
\hline
 \end{tabular}
\caption{ISCO radius $r_{ISCO}$, OSCO radius $r_{OSCO}$, specific angular momentum $(\Lb_{ISCO}, \Lb_{OSCO})$ and specific energy $(\mathscr{E}_{ISCO}, \mathscr{E}_{OSCO})$ with varying $\eta$ and $L$.}
    \label{tab SCO}
\end{table}

\begin{figure}
\centering

\includegraphics[width=200pt,height=160pt]{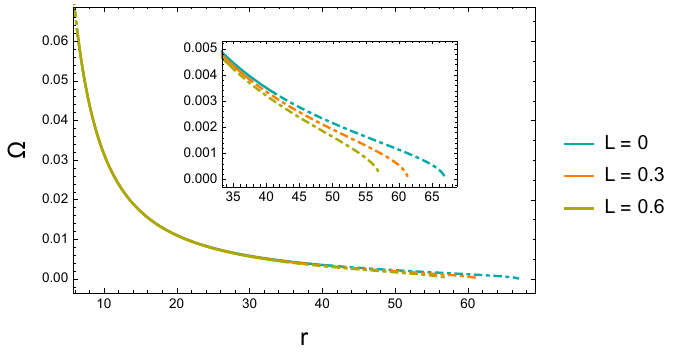}
\caption{Behaviour of the Keplerian frequency for various choices of $L$ with $\lambda=10^{-5}$, $M=1$, $\Lb=5$ and $\eta=0.05$. The solid and dotdashed portions of the curve correspond to the positions of stable and unstable circular orbits, respectively.}
\label{fig Keplefre}
 \end{figure}

The orbital angular velocity of massive particles moving in circular orbits is given by \cite{Ahmad2025}
\begin{eqnarray}\label{eqn Keplefre}
\Omega=\dfrac{\dot{\phi}}{\dot{t}}=\sqrt{\dfrac{f'(r)}{2 r}}=\sqrt{\dfrac{3M-(1+L)\Lambda r^3}{3 r^3}}.
\end{eqnarray}
This angular velocity is known as the Keplerian frequency. It can be seen that the Keplerian frequency shows dependence on the LV parameter $L$, the cosmological constant $\Lambda$. However, the frequency is independent of the GM parameter, which is evident from the definition of the angular velocity and the expression of $f(r)$. Fig. \ref{fig Keplefre} shows the radial dependence of the Keplerian frequency for massive particles moving around the BH. It is found that the frequency monotonically decreases with $r$ and the frequency is lowered with a larger LV parameter. We also identify that larger $L$ values result in the Keplerian frequency becoming zero at a smaller radial distance.\\

\begin{figure}[h!]
\centering
  \subfloat[\centering ]{{\includegraphics[width=170pt,height=160pt]{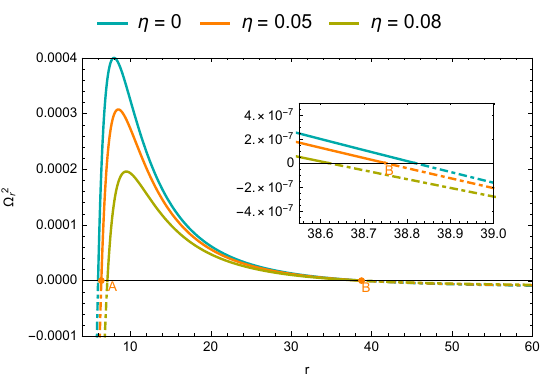}}\label{fig RadialepiEta}}
  \qquad
   \subfloat[\centering ]{{\includegraphics[width=170pt,height=160pt]{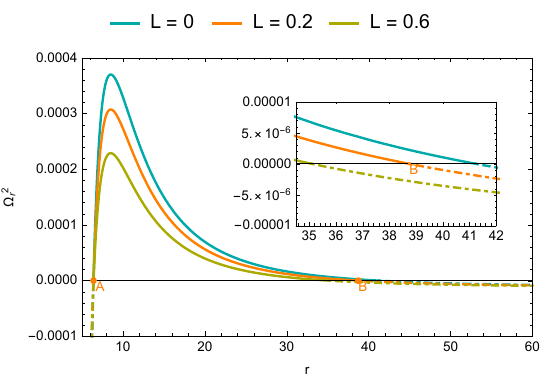}}\label{fig RadialepiL}}
   \caption{Behaviour of the radial epicyclic frequency for different choices of  $\eta$ (a) and $L$ (b) with $M=1$, $\Lb=5$, $\Lambda=10^{-5}$. Here,(a) $L=0.2$ and (b) $\eta=0.05$.}
   \label{fig Radialepi}
\end{figure}

\begin{figure}
\centering

\includegraphics[width=200pt,height=160pt]{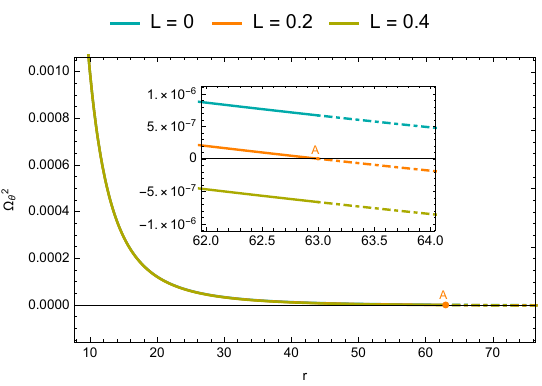}
\caption{Behaviour of the vertical epicyclic frequency for different choices of $L$ with $M=1$, $\Lambda=10^{-5}$.}
\label{fig vertEpiFre}
 \end{figure}

We will examine the epicyclic frequencies to study the orbit properties of massive particles. We will investigate the vertical epicyclic frequency $\Omega_{\theta}^2$ and the radial epicyclic frequency $\Omega_r^2$ of the particles to establish the radial positions where the equatorial circular motion is unstable or stable in the vertical and radial directions. Eqs. \eqref{eqn components} and \eqref{eqn radial} can we rewritten as 
\begin{eqnarray}\label{eqn epi fre r}
\dfrac{1}{2}\bigg(\dfrac{dr}{dt}\bigg)^2= -\dfrac{1}{2}\dfrac{f(r)^3}{E^2}\bigg(\dfrac{-E^2}{(1+L)f(r)}+ \dfrac{1}{1+L}+\dfrac{L^2}{(1+L)r^2\sin^2\theta}\bigg)\equiv V^{(r)},\\\label{eqn epi fre theta}
\dfrac{1}{2}\bigg(\dfrac{d\theta}{dt}\bigg)^2= -\dfrac{1}{2}\dfrac{f(r)^2}{r^2 E^2}\bigg(\dfrac{-E^2}{f(r)}+ 1+\dfrac{L^2}{r^2\sin^2\theta}\bigg)\equiv V^{(\theta)}.
\end{eqnarray}
Here, $\sin^2\theta$ is retrieved to examine the orbital perturbations. For an equatorial circular orbit, the above two equations govern the radial and vertical motions. We take the coordinate time derivative of Eqs. \eqref{eqn epi fre r} and \eqref{eqn epi fre theta} by introducing small displacements $\delta r$ and $\delta \theta$ to give \cite{Xiang-Qian 2025} 
\begin{eqnarray}
\dfrac{d^2 \delta r}{dt^2}=\dfrac{d^2 V^{(r)}}{dr^2}\delta r,  \hspace{0.8 cm}  \dfrac{d^2 \delta \theta}{dt^2}=\dfrac{d^2 V^{(\theta)}}{d \theta^2}\delta \theta.
\end{eqnarray}
The epicyclic frequencies can be evaluated from the above two equations. The radial frequency and the vertical epicyclic frequency can be deduced as
\begin{eqnarray}
\Omega_r^2=\dfrac{d^2 V^{(r)}}{dr^2}, \hspace{0.8 cm} \Omega_{\theta}^2=\dfrac{d^2 V^{(\theta)}}{d \theta^2}.
\end{eqnarray}
Using Eqs. \eqref{eqn L2}, \eqref{eqn epi fre r} and \eqref{eqn epi fre theta}, the epicyclic frequencies can be expressed as 
\begin{eqnarray}
\Omega_r^2&=&\dfrac{1}{2(1+L)}\bigg({\dfrac{3 f'(r) f(r)}{r}+ f''(r) f(r)-2 f'(r)^2}\bigg),\\
\Omega_{\theta}^2&=&\dfrac{f'(r)}{2 r}=\Omega_K,
\end{eqnarray} 
where $\Omega_K$ is the Keplerian frequency. The radial profile of $\Omega_r^2$ for particles in circular stable orbits in a Schwarzchild-dS-like BH spcatime with GM in Bumblebee gravity is presented in Fig. \ref{fig Radialepi} for various choices of $L$ and $\eta$. From the figure, we can see that the incorporation of $\eta$ and $L$ reduces the maximum radial epicyclic frequency. The region where $\Omega_r^2$ is positive indicates the interval where the circular equatorial orbits have radial stability. In the Figs. \ref{fig RadialepiEta} and \ref{fig RadialepiL}, we observe that for $\eta=0.05$ and $L=0.2$, $\Omega_r^2$ is positive in the interval between $r_A$ and $r_B$ indicating stability of the radial epicyclic frequency. The points $r_A$ and $r_B$ correspond to the points $r_{ISCO}$ and $r_{OSCO}$ in Fig. \ref{fig TimeVeffL} respectively. In Fig. \ref{fig vertEpiFre}, we illustrate the radial profile of the vertical epicyclic frequency $\Omega_{\theta}^2$ for different choices of $L$. From the figure, we see that the circular orbits are stable vertically as well, in the range where $\Omega_r^2$ is stable, as demonstrated by the positive values of $\Omega_{\theta}^2$ before dropping to zero at radial distances greater than the stable range of $\Omega_r^2$. In particular, for $L=0.2$ $\Omega_{\theta}^2$ becomes zero at $r_A$, which corresponds to the point J in Fig. \ref{fig TimeVeffL}.   

\begin{figure}[h!]
\centering
  \subfloat[\centering ]{{\includegraphics[width=130pt,height=130pt]{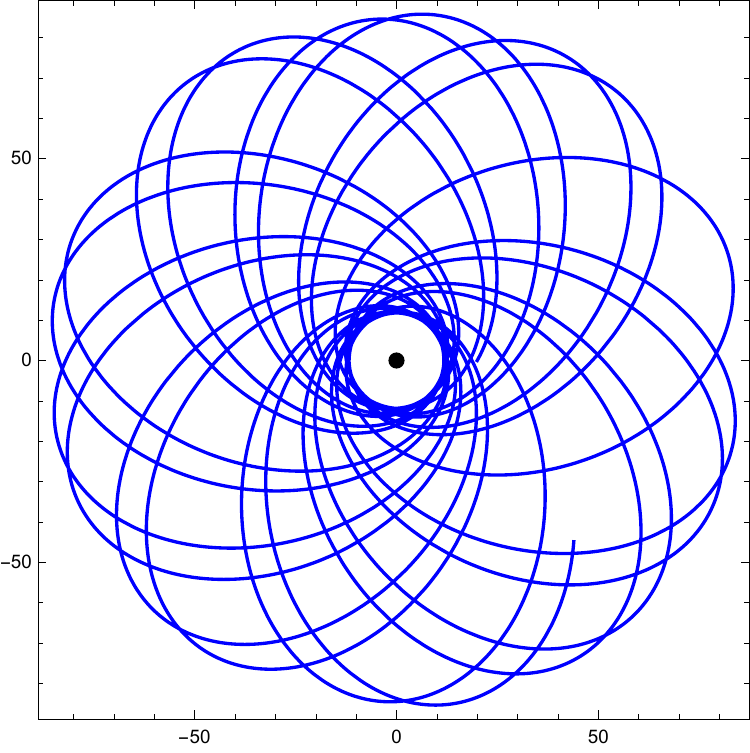}}\label{fig TTraEta0}}
  \quad
   \subfloat[\centering ]{{\includegraphics[width=130pt,height=130pt]{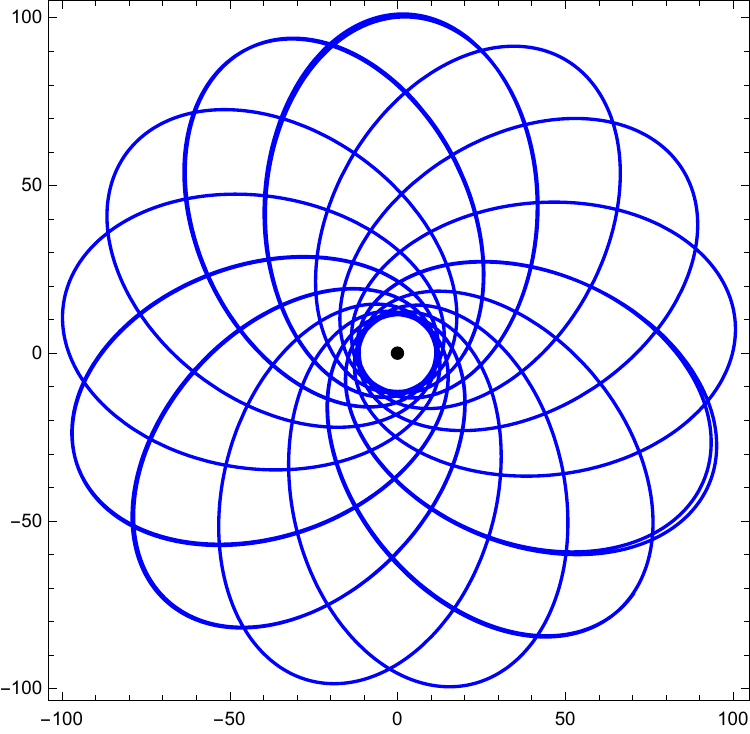}}\label{fig TTraEta01}}
   \quad
   \subfloat[\centering ]{{\includegraphics[width=130pt,height=130pt]{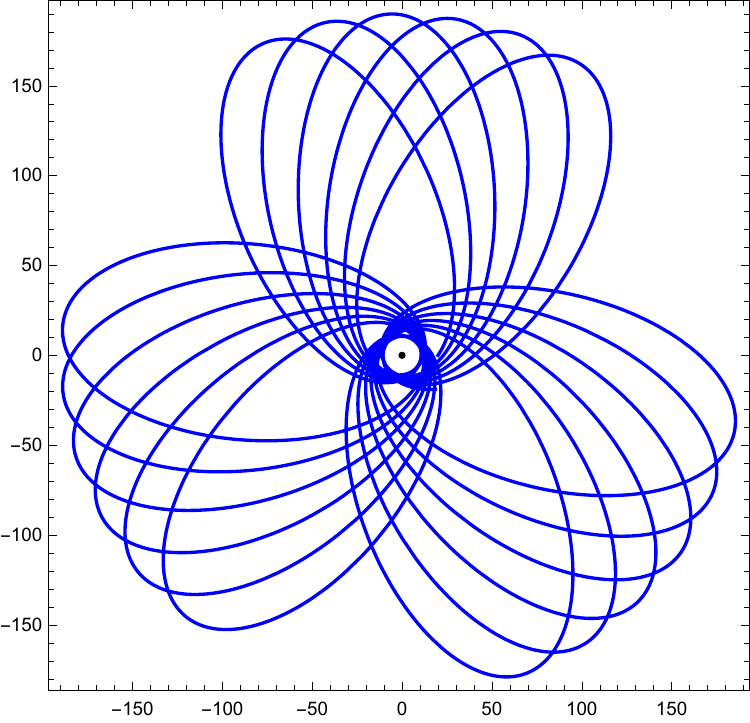}}\label{fig TTraEta02}}
   \caption{Illustration of timelike particle trajectories varying $\eta$ with  $\Lb=5$, $L=0.2$, $\Lambda=10^{-10}$ and $M=1$. Here, (a) $\eta=0$ (b) $\eta=0.01$ (c) $\eta=0.02$.}
   \label{fig TimetraEta}
\end{figure}

\begin{figure}[h!]
\centering
  \subfloat[\centering ]{{\includegraphics[width=130pt,height=130pt]{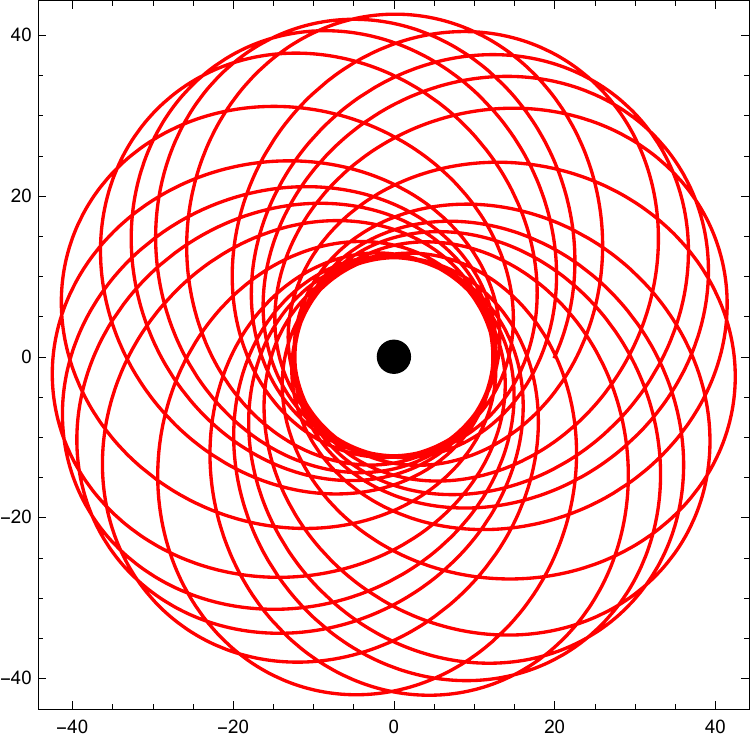}}\label{fig TTra0}}
  \quad
   \subfloat[\centering ]{{\includegraphics[width=130pt,height=130pt]{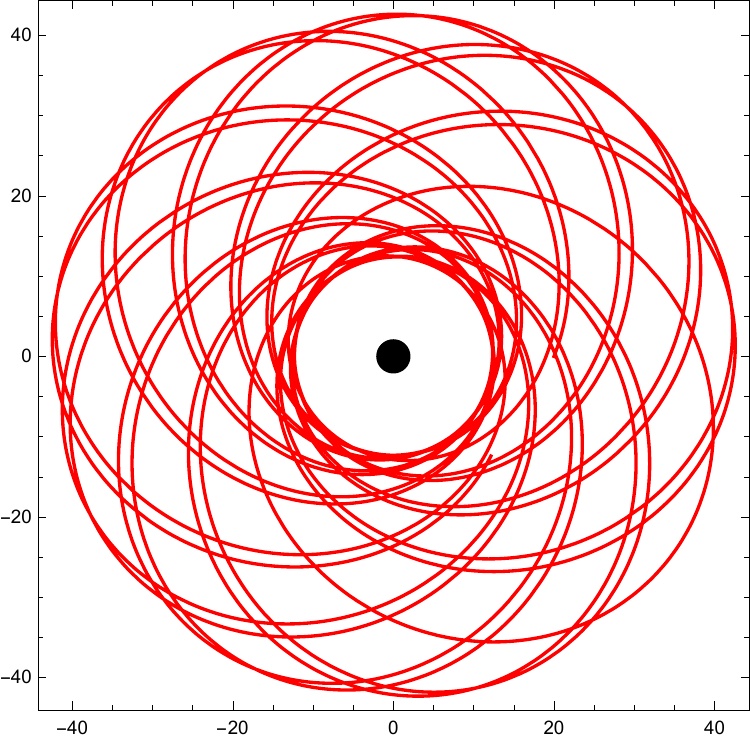}}\label{fig TTraL03}}
   \quad
   \subfloat[\centering ]{{\includegraphics[width=130pt,height=130pt]{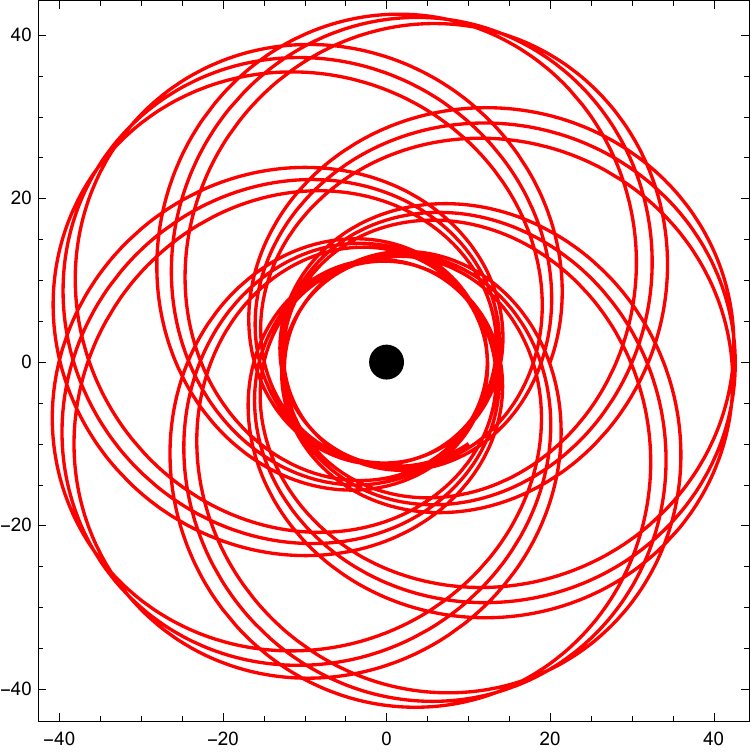}}\label{fig TTraL06}}
   \caption{Illustration of timelike particle trajectories varying $L$ with  $\Lb=5$, $\Lambda=10^{-10}$, $\eta=0.05$ and $M=1$. Here, $L=0, 0.3$ and $0.6$ in panels (a), (b) and (c) respectively.}
   \label{fig TimetraL}
\end{figure}

\begin{figure}[h!]
\centering
  \subfloat[\centering ]{{\includegraphics[width=130pt,height=130pt]{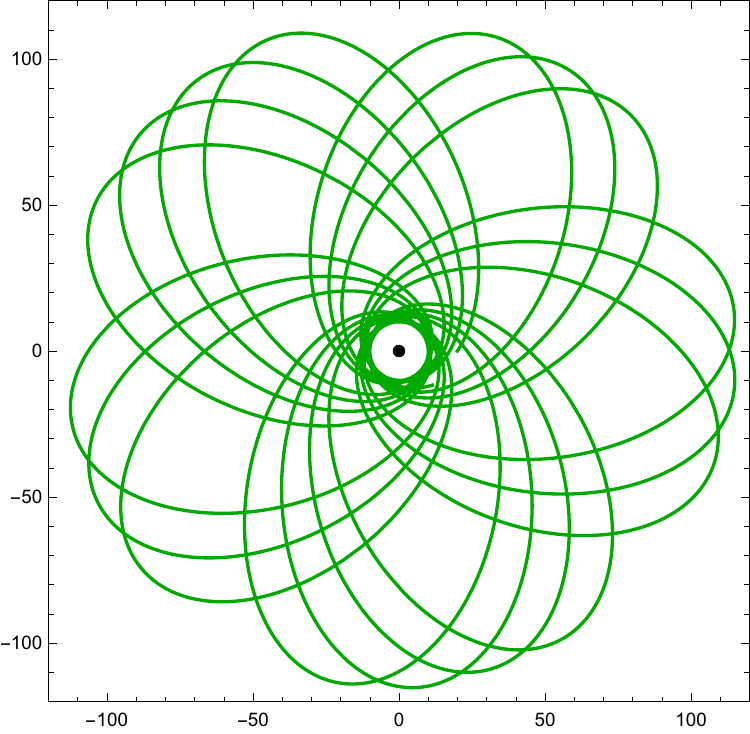}}\label{fig TTraL15}}
  \quad
   \subfloat[\centering ]{{\includegraphics[width=130pt,height=130pt]{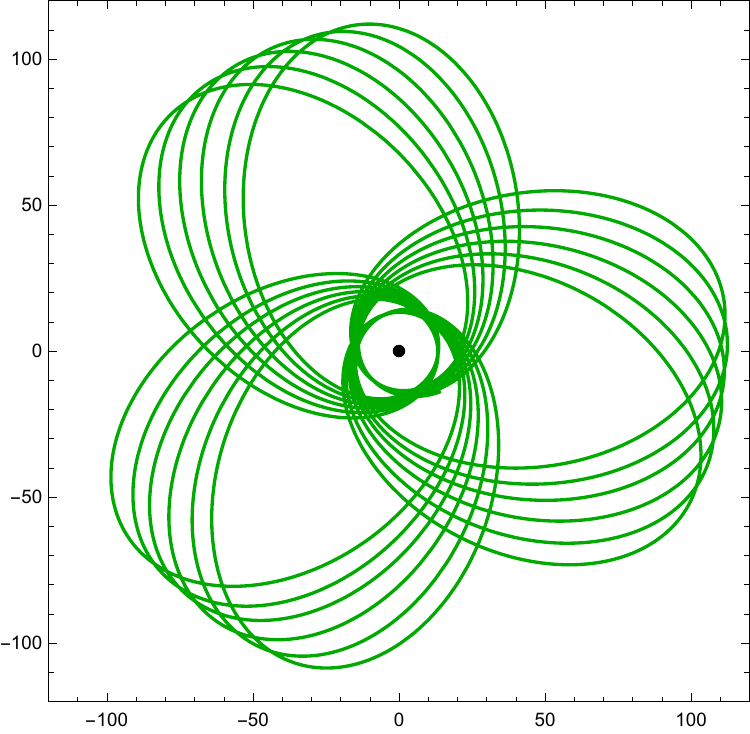}}\label{fig TTraL155}}
   \quad
   \subfloat[\centering ]{{\includegraphics[width=130pt,height=130pt]{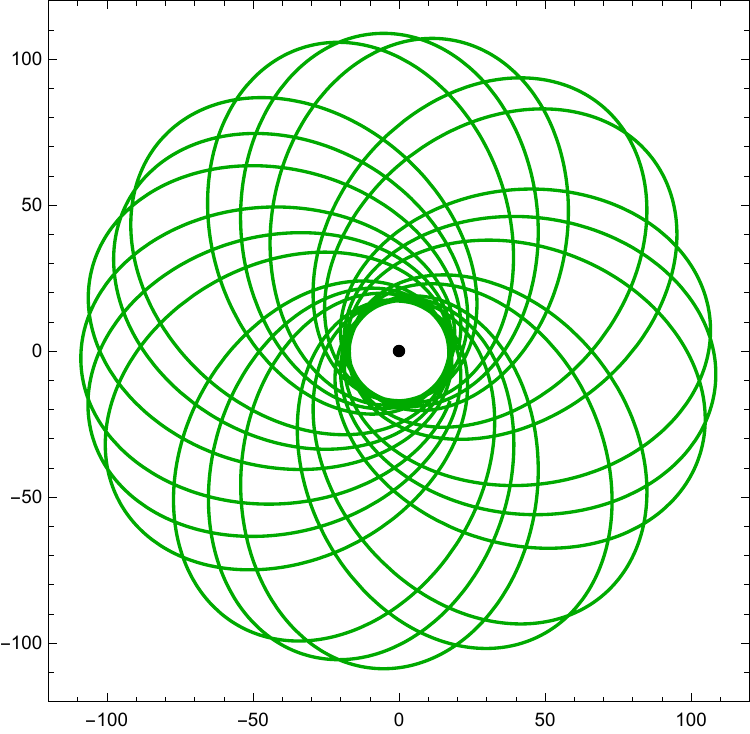}}\label{fig TTraL16}}
   \caption{Illustration of timelike particle trajectories varying $\Lb$ with  $L=0.2$, $\Lambda=10^{-10}$, $\eta=0.05$ and $M=1$. Here, (a) $\Lb=5$ (b) $\Lb=5.5$ (c) $\Lb=6$.}
   \label{fig TimetraL1}
\end{figure}

The equation of motion governing the trajectory of timelike particles in the spacetime is given by 
\begin{eqnarray}
\bigg( \dfrac{1}{r^2}\dfrac{dr}{d \phi}\bigg)^2=\dfrac{\mathscr{E}}{\Lb^2}-\bigg(\dfrac{1}{\Lb^2}+\dfrac{1}{r^2}\bigg)\dfrac{1}{(1+L)}\bigg(1-\kappa \eta^2 -\dfrac{2M}{r}-\dfrac{(1+L)\Lambda r^2}{3}\bigg).
\end{eqnarray}
Making the transformation $r=\dfrac{1}{u}$, we have 
\begin{eqnarray}\label{eqn time tra}
\bigg(\dfrac{du}{d \phi}\bigg)^2=\dfrac{\mathscr{E}}{\Lb^2}-\dfrac{1}{(1+L)}\bigg[\dfrac{1}{\Lb^2}-\dfrac{\kappa \eta^2}{\Lb^2}-\dfrac{2M u}{\Lb^2}-\dfrac{(1+L)\Lambda}{3 u^2 \Lb^2}+u^2-u^2 \kappa \eta^2- 2 M u^3-\dfrac{(1+L)\Lambda}{3}\bigg].
\end{eqnarray}
Differentiating Eq. \eqref{eqn time tra} w.r.t. $\phi$, we get a second order nonlinear differential equation
\begin{eqnarray}\label{eqn timeliketra}
\dfrac{d^2u}{d\phi^2}+\bigg(\dfrac{1}{1+L}-\dfrac{\kappa \eta^2}{1+L}\bigg)u=\dfrac{M}{\Lb^2 (1+L)}-\dfrac{\Lambda}{3 \Lb^2 u^3}+\dfrac{3 M u^2}{1+L}.
\end{eqnarray}
Eq. \eqref{eqn timeliketra} depicts the trajectory of timelike particles in the gravitational field of the spacetime. From the above equation, it is evident that the trajectory of massive particles in the spacetime is dependent on factors including the GM parameter $\eta$, the LV parameter $L$, the cosmological constant $\Lambda$ and the angular momentum $\Lb$. Figs. \ref{fig TimetraEta}, \ref{fig TimetraL} and  \ref{fig TimetraL1} demonstrate the relativistic orbital dynamics for massive particles in the the Schwarzschild-dS-like BH spacetime with GM. We have shown the behaviour of the trajectories for varying $\eta$, $L$ and the angular momentum $\Lb$. Here, we limit our observation to the changes brought about by the parameters on the shape of the orbits qualitatively.  From Fig. \ref{fig TimetraEta}, it is observed that as $\eta$ increases, the orbital structure undergoes a systematic expansion, where the apastron broadens and the periastron contracts. The orbits are farther from the BH in the absence of the GM parameter. The turning points of the effective potential are the positions where $\dot{r}$ equals zero momentarily so that the motion is oscillatory and repeats in that direction \cite{Mrinnoy M 2025}. These turning points of the effective potential correspond to the apastron and periastron, which are the points $r_{max}$ and $r_{min}$. We also notice that the rosette patterns become denser with increasing $\eta$ with enhanced orbital precession. This suggests that the effective gravitational field is weaker at higher $\eta$ values. Furthermore, we observe that small changes in the value of $\eta$ significantly affect the orbital structure of massive particles. Therefore, the massive particle orbits are sensitive to small variations in the choice of the GM parameter in the spacetime.  From Fig. \ref{fig TimetraL}, we observe a noticeable difference in the orbital structure characterized by a wider orbital pattern, with the rise of the LV parameter $L$. Significant variations in the orbital pattern are also observed when the angular momentum $\Lb$ is varied as shown in Fig. \ref{fig TimetraL1}. It is also observed that the orbits are shifted to larger radial distances from the BH with the rise of the angular momentum value. We also observe that the orbits show smaller radial excursions with higher $\Lb$ and display a contraction of the orbital structure.

\begin{figure}[h!]
\centering
  \subfloat[\centering ]{{\includegraphics[width=170pt,height=160pt]{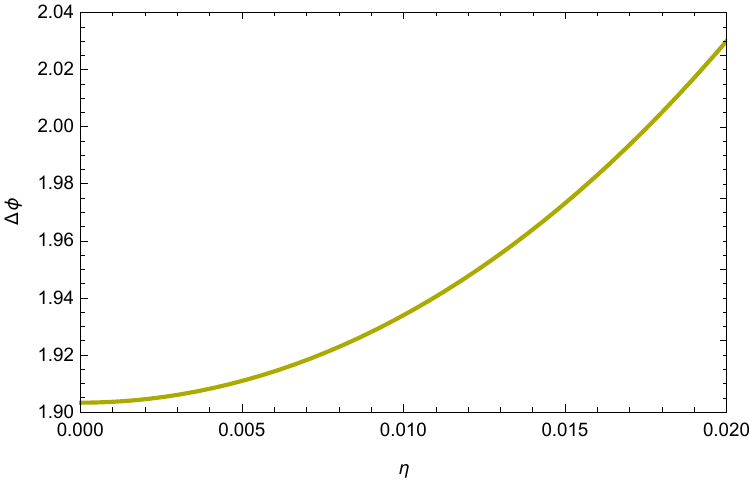}}\label{fig Precessioneta}}
  \qquad
   \subfloat[\centering ]{{\includegraphics[width=170pt,height=160pt]{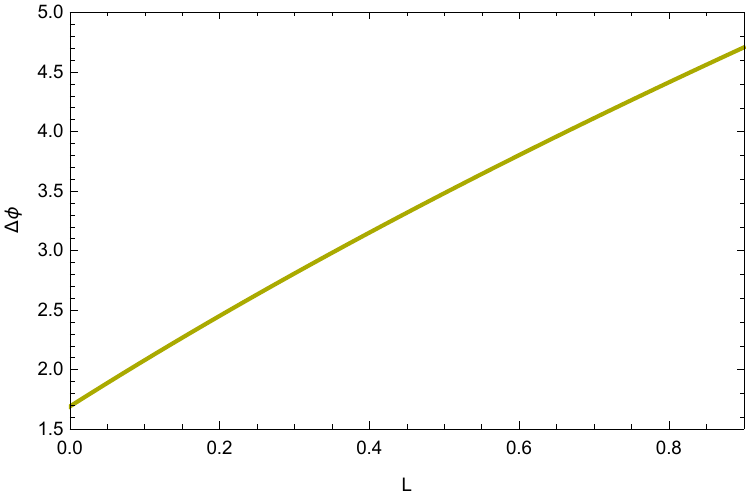}}\label{fig PrecessionL}}
   \caption{Variation of $\Delta \phi$ w.r.t. $\eta$(a) and $L$ (b) with $M=1$, $\Lb=5$, $\Lambda=10^{-5}$. Here,(a) $L=0.2$ and (b) $\eta=0.05$.}
   \label{fig Precession}
\end{figure}

Now, we will investigate the perihelion precession for orbits of timelike particles in the Schwarzschild-dS like BH spacetime with GM. Eq. \eqref{eqn lagran} can be written as 
\begin{eqnarray}
\bigg( \dfrac{dr}{d \phi}\bigg)^2=\dfrac{\mathscr{ E}r^4}{\Lb^2}-\bigg(\dfrac{1}{\Lb^2}+\dfrac{1}{r^2}\bigg)\dfrac{r^4}{(1+L)}f(r).
\end{eqnarray}  
$\dfrac{dr}{d \phi}=0$ at the perihelion and the aphelion \cite{Mohammas 2022}. The precession of the azimuthal angle $\phi$ for massive particle orbits can be calculated as \cite{Mrinnoy M 2025}
\begin{eqnarray}
\Delta\phi&=&\int_{r_{min}}^{r_{max}}\dfrac{d\phi}{dr}dr+\int_{r_{max}}^{r_{min}}\dfrac{d\phi}{dr}dr-2 \pi\nonumber\\
&=&2\int_{r_{min}}^{r_{max}}\dfrac{dr}{\dfrac{r^2}{\sqrt{1+L}}\sqrt{\dfrac{E^2}{\Lb^2}-\dfrac{f(r)}{\Lb^2}-\dfrac{f(r)}{r^2}}}-2 \pi,
\end{eqnarray}
where $r_{min}$ and $r_{max}$ are the minimal and maximal distance of the timelike orbit from the BH. Fig. \ref{fig Precession} displays the behaviour of the precession angle for varying $\eta$ and $L$. It is observed that the precession angle exhibits a monotonic increase with the rise of $\eta$ and the rate of increase is enhanced for higher values of $\eta$. In addition, it is evident that the precession angle is also enhanced with $L$ and follows a roughly linear trend.  

\subsection{Lyapunov Stability}
\begin{figure}[h!]
\centering
  \subfloat[\centering ]{{\includegraphics[width=170pt,height=160pt]{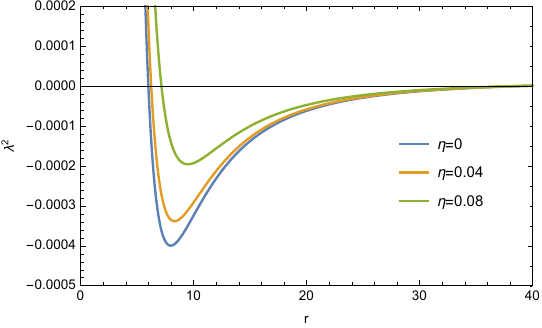}}\label{fig TimeLyaeta}}
  \qquad
   \subfloat[\centering ]{{\includegraphics[width=200pt,height=160pt]{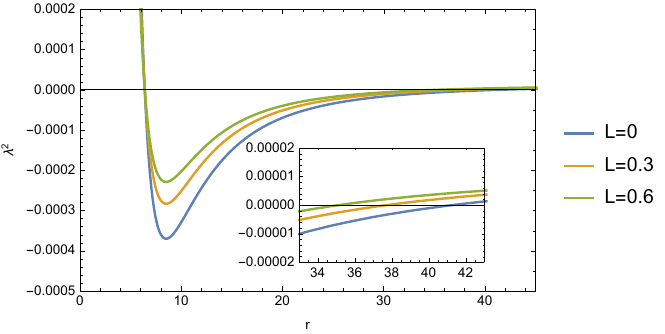}}\label{fig TimeLyaL}}
   \caption{Variation of $\lambda^2$ w.r.t.  $\eta$ (a) and $L$ (b) with $M=1$, $\Lb=5$, $\Lambda=10^{-5}$. Here,(a) $L=0.2$ and (b) $\eta=0.05$.}
   \label{fig TimeLya}
\end{figure}

 The stability analysis of the circular orbits of massive particles is crucial in the study of particle dynamics around a black hole. The Lyapunov exponent is a tool to investigate the trajectory behaviour in orbital motion \cite{Javlon2024}. From Eq. \eqref{eqn Lya}, we have the expression for Lyapunov exponent as 
\begin{eqnarray}\label{eqn Lya2}
\lambda=\sqrt{-\dfrac{V_{eff}''(r)}{2 \dot{t}^2}},
\end{eqnarray}
where $\dot{t}=\dfrac{E}{f(r)}$. Using the expression of $f(r)$ and Eq. \eqref{eqn Veff time}, we derive the expression of the Lyapunov exponent as
\begin{eqnarray}\label{eqn Lya time} 
\lambda^2=\dfrac{18M^2-4(1+L)r^4(-1+8 \pi \eta^2)\Lambda-3 M(r-8 \pi r \eta^2+5(1+L)r^3\Lambda)}{3 (1+L)r^4}.
\end{eqnarray}
 In Fig.\ref{fig TimeLya}, the radial dependence of the Lyapunov exponent is displayed for different values of $\eta$ and $L$. The Lyapunov exponent curves cross the radial coordinate axis at two points. The smaller intersection point corresponds to the ISCO radius of the associated curve, and the larger intersection point corresponds to the OSCO radius of each curve. It is observed from Fig. \ref{fig TimeLyaeta} that increasing $\eta$ enhances the ISCO radius slightly while the stable orbit existing region shrinks. Also from Fig.\ref{fig TimeLyaL}, we notice that higher $L$ values lead to a reduction in the OSCO radius, thereby shrinking the stable orbit existing region. The narrowing effect of the stable region is more significant in the case of increasing $L$. Beyond the ISCO and OSCO radii, the Lyapunov exponent is positive, which indicates instability of the orbits in the region. The region between the ISCO and OSCO radii is negative, indicating stability in the region. Additionally, we observe that the stability of the orbits of timelike particles diminishes as $\eta$ and $L$ increase. 

\section{Eikonal QNMs}

It is known that a connection exists between the unstable null geodesics and QNMs of BH in the eikonal limit. The characteristics of unstable null circular orbits are directly associated with the eikonal $(l>>1)$ 
quasinormal spectrum of a BH \cite{Mrinnoy Ma 2025}. This suggests a link between the possible trajectories of light rays and the BH's response to external perturbations. The eikonal-limit effective potential for scalar perturbation can be described as \cite{M.S. Churilova}:
\begin{eqnarray}
V_{eik}(r)=f(r)\bigg(\dfrac{l(l+1)}{r^2}\bigg).
\end{eqnarray}
 This implies that the electromagnetic and scalar perturbations are governed by equivalent effective potentials in the eikonal limit.for static spacetimes \cite{Saraswati 2020}. For spherically symmetric, static and asymptotically flat spacetimes, the multiples of the Lyapunov exponent and angular frequency of the circular unstable null geodesics give the imaginary and real parts of the QNMs, in the eikonal approximation \cite{Vitor 2009}. The real part of the QNMs corresponds to the product of the $l$ with $\Omega_{ph}$, the angular frequency of the orbiting unstable photons, while the imaginary part corresponds to the product of $(n+1/2)$ and the Lyapunov exponent at the unstable null geodesic. It is written as \cite{Shobhit 2022}
\begin{eqnarray}
\omega_{QNM}=l \Omega_{ph}-i\bigg(n+\dfrac{1}{2}\bigg)\lambda,
\end{eqnarray}
where $n$,  $l>>1$ represent the overtone number and the angular momentum number, respectively and $\Omega_{ph}=\sqrt{f(r_{ph})}/r_{ph}$.
Using Eq. \eqref{eqn Lyanull}, we get the expression of the QNMs in the eikonal limit of the spacetime as 
\begin{eqnarray}\label{eqn QNM}
\omega_{QNM}=l \dfrac{\sqrt{f(r_{ph})}}{r_{ph}}-i \bigg(n+\dfrac{1}{2}\bigg)\sqrt{\dfrac{(-1+8 \pi \eta^2)(6M +r_{ph}(-3+24 \pi \eta^2+r_{ph}^2 \Lambda+L r_{ph}^2 \Lambda))}{3(1+L)r_{ph}^3}}.
\end{eqnarray}
From Eq. \eqref{eqn QNM}, we observe that the QNMs depend on the GM parameter $\eta$, the LV parameter $L$, the cosmological constant $\Lambda$ and the radius of the unstable circular null geodesic $r_{ph}$. Also, we see that the real part of the QNMs does not depend on the overtone number $n$ and the imaginary part of the QNMs is independent of the angular momentum number $l$.

The scalar, Dirac and electromagnetic perturbations for the spacetime was investigated in \cite{YenshembamNuclear2025} and the associated QNMs was also computed. Using the expressions of the effective potentials derived in \cite{YenshembamNuclear2025}, we compute the quasinormal frequencies in the large multipole $\ell$ regime through the WKB method and compare it with the QNMs obtained from \eqref{eqn QNM}. This comparison enables us to explore the correspondence between the unstable null geodesics and the QNMs of black holes in the eikonal limit. The agreement between these two approaches will demonstrate the extent to which the quasinormal spectrum is characterized by the underlying null-geodesic structure of the spacetime.\\
 Tables \ref{tab_compare_e} and \ref{tab_compare_L} provide the comparison between the QNMs for large choices of $\ell$. We consider various choices of $\eta$ and $L$ to observe the changes induced by the GM and the LV parameters on the QNM frequencies. The real and imaginary component of the QNMs are associated with the oscillation frequency and damping timescale which can be used to examine the stability of the spacetime. From Table \ref{tab_compare_e}, it is evident that the real part of the QNMs are reduced as $\eta$ increases and the decrease becomes steeper with larger $\eta$. Meanwhile, the imaginary part also decreases with the rise of $\eta$ and the decline rate gradually increases with $\eta$. This implies that higher monopole parameter results in a lower oscillation frequency and damping rate, indicating that larger values of the GM parameter contribute to greater black hole stability. Moreover, we see that the QNMs for the scalar and electromagnetic perturbations are the same and the imaginary part of the QNMs for both the perturbations closely match the QNMs computed using the eikonal approximation formula. And, the real part of the QNMs in the eikonal limit formula differs slightly from the first two columns. The relative error differs only marginally with values ranging from 0.000484001 to 0.000498857, displaying minor variations with $\eta$. Similarly, from Table \ref{tab_compare_L}, we notice that the imaginary part of the QNMs agrees closely for both methods, while the real part shows minor variation after the decimal point. Furthermore, we notice that changes in the real and imaginary components of the QNMs due to increase in $L$ is small but it is consistent, revealing that the QNMs are mildly influenced by variations in $L$. Increasing $L$ produce a slight reduction in both the damping rate and oscillation frequency. Moreover, we notice that the relative error is almost the same for varying $L$.   The above observations indicate that the results from the two methods align closely with each other.

\begin{figure}[h!]
\centering
  \subfloat[\centering ]{{\includegraphics[width=170pt,height=160pt]{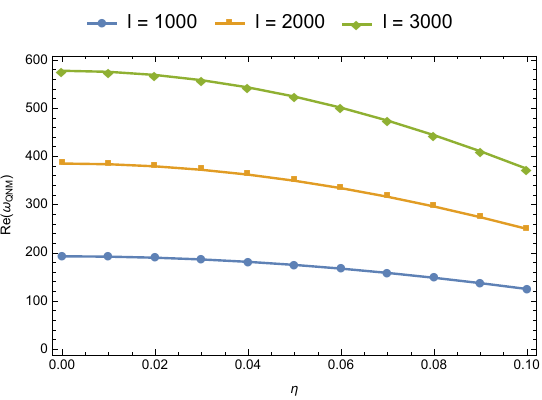}}\label{fig QNMEta}}
  \qquad
   \subfloat[\centering ]{{\includegraphics[width=170pt,height=160pt]{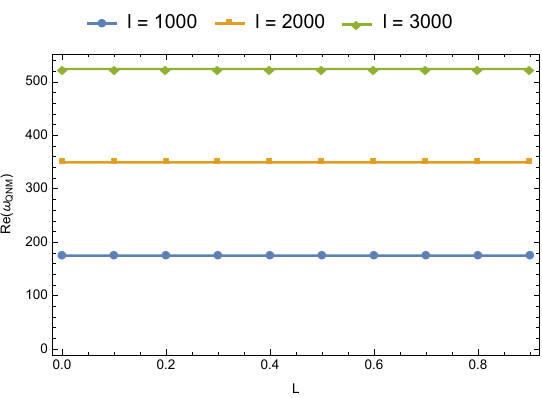}}\label{fig QNML}}
   \caption{Plot of the real part of the fundamental eikonal QNMs as a function of  $\eta$ (left) and $L$ (right) for different choices of $\ell$ with  $M=1$, $\Lambda=10^{-5}$. Here, (a) $L=0.2$ and (b) $\eta=0.05$.}
   \label{fig QNM}
\end{figure}

In Fig. \ref{fig QNM}, we show the dependence of the real part of the fundamental eikonal QNMs on $\eta$ and $L$ for different choices of $\ell$. We notice that $\eta$ has a much larger effect on the real part of the QNMs as compared to $L$. The real part of the QNMs decreases monotonically as $\eta$ becomes larger, and the QNMs increase with larger $\ell$ values as seen from the figure.

\begin{table}[h]
\centering
\begin{tabular}{c c c c c}
\toprule
$\eta$ &
Scalar  &
Electromagnetic &
Eikonal & $\vert  \Delta \omega / \omega \vert$\\
\midrule
0    & $192.536 - 0.0878363i$ & $192.536 - 0.0878363i$ & $192.44-0.08784i$ &$0.000498857$ \\
0.03 & $186.04 - 0.0839073i$ & $186.04 - 0.0839073i$  & $185.95-0.08391i$ &$0.000484001$ \\
0.06 & $167.003 - 0.0726596 i$ & $167.003 - 0.0726596 i$ & $166.92-0.07266i$  &$0.000497244$ \\
0.09 & $136.838 - 0.055711i$ & $136.838 - 0.055711i$ & $136.77-0.05571i$ &$0.000497185$\\
\bottomrule
\end{tabular}
\caption{ Comparison of scalar, electromagnetic and eikonal quasinormal frequencies for various choices of $L$ for the mode $n = 0$  and $\ell=1000$ with fixed at $M=1$, $\Lambda=10^{-5}$ and $\L=0.2$.}
\label{tab_compare_e}
\end{table}

\begin{table}[h]
\centering
\begin{tabular}{c c c c c}
\toprule
$L$ &
Scalar &
Electromagnetic  &
Eikonal   & $\vert  \Delta \omega / \omega \vert$ \\
\midrule
0    & $	174.678 - 0.0845083	i$ & $174.678 - 0.0845083i$ & $	 174.591-0.0845083 i	$ &$0.000498307$ \\
0.2 & $	174.676 - 0.0771443	i$ & $174.676 - 0.0771443 i $ & $	 174.589-0.0771443i$  &$0.000498313$ \\
0.4 & $	174.674 - 0.071421	 i$ & $	174.674 - 0.071421	 i$ & $ 174.587-0.071421 i	$   &$0.000498319$ \\
0.6 & $	174.672 - 0.0668075	i$ &  $	174.672 - 0.0668075	i$ & $	 174.585-0.0668075 	i$  &$0.000498325$\\
\bottomrule
\end{tabular}
\caption{ Comparison of scalar, electromagnetic and eikonal quasinormal frequencies for various choices of $L$ for the mode $n = 0$  and $\ell=1000$ with fixed at $M=1$, $\Lambda=10^{-5}$ and $\eta=0.05$.}
\label{tab_compare_L}
\end{table}

\section{Extended Phase Space Thermodynamics and Holographic Heat Engine}

In the framework of extended phase space thermodynamics, the cosmological constant $\Lambda$ is no longer treated as a fixed geometric parameter but is promoted to a dynamical thermodynamic pressure $P$. For asymptotically Anti-de Sitter (AdS) black holes in Bumblebee gravity, the thermodynamic pressure is modified by the Lorentz-violating parameter $L$, defined as \cite{EslamPanah2026}:
\begin{eqnarray}\label{eqn pressure}
P = -\dfrac{\Lambda(1+L)}{8\pi}.
\end{eqnarray}
Correspondingly, the mass of the black hole is interpreted as the chemical enthalpy of the spacetime, $\mathcal{M} \equiv M_{AMD}$. Using the Ashtekar-Magnon-Das (AMD) approach for Bumblebee backgrounds, the physical mass $\mathcal{M}$ is scaled by the Lorentz violation parameter as $\mathcal{M} = \frac{M}{\sqrt{1+L}}$ \cite{EslamPanah2026}. By evaluating the lapse function $f(r_h) = 0$ at the event horizon and substituting Eq. \eqref{eqn pressure}, the enthalpy is expressed as:
\begin{eqnarray}\label{eqn enthalpy}
\mathcal{M} = \dfrac{r_h(1-\kappa\eta^2)}{2\sqrt{1+L}} + \dfrac{4\pi P r_h^3}{3\sqrt{1+L}}.
\end{eqnarray}

The Hawking temperature $T$ of the black hole is obtained from the surface gravity $\kappa_{sg}$, yielding:
\begin{eqnarray}\label{eqn temp}
T = \dfrac{f'(r_h)}{4\pi \sqrt{1+L}} = \dfrac{1-\kappa\eta^2}{4\pi r_h \sqrt{1+L}} + \dfrac{2P r_h}{\sqrt{1+L}}.
\end{eqnarray}
By treating $S$, $P$, $L$, and $\eta$ as independent state variables, the conjugate thermodynamic volume $V$ is given by the equation of state $V = \left( \partial \mathcal{M} / \partial P \right)_{S, L, \eta} = 4\pi r_h^3 / (3\sqrt{1+L})$.

\subsection{Mathematical Origin of the Modified Entropy}
Before evaluating the thermal stability and heat engine efficiency, we must rigorously address the entropy of the spacetime. Under the classical Bekenstein-Hawking formulation, the entropy follows the standard area law $S_0 = \pi r_h^2$. However, in the high-energy regimes where spontaneous Lorentz symmetry breaking occurs, semiclassical entropy inevitably receives quantum corrections. 

Mathematically, thermal fluctuations around equilibrium states and loop quantum gravity microstate counting both yield a leading-order logarithmic correction to the classical entropy, typically of the form $S_{mod} = S_0 - k \ln(S_0)$. Because the Bumblebee vector field dynamically alters the phase space volume and the dispersion relations of microstates near the horizon, the correction prefactor $k$ must be intimately coupled to the Lorentz-violating parameter $L$. Therefore, we propose a physically consistent modified entropy for the Bumblebee framework as:
\begin{eqnarray}\label{eqn mod_entropy}
S_{mod} = \pi r_h^2 - \alpha L \ln(\pi r_h^2),
\end{eqnarray}
where $\alpha$ is a dimensionless positive constant tracking the strength of the microstate correction induced by the Lorentz violation. The differential of this modified entropy is straightforwardly evaluated as $dS_{mod} = 2(\pi r_h - \frac{\alpha L}{r_h})dr_h$.

\subsection{Local Thermal Stability and Corrected Specific Heat}
The local thermodynamic stability of the black hole is governed by the specific heat at constant pressure, $C_P = T (\partial S / \partial T)_P$. A positive $C_P$ signifies a thermodynamically stable phase, whereas a negative $C_P$ indicates a state of instability. Using the classical entropy $S_0$, the specific heat evaluates to:
\begin{eqnarray}\label{eqn cp_class}
C_{P, 0} = \dfrac{2\pi r_h^2 \left( 1 - \kappa\eta^2 + 8\pi P r_h^2 \right)}{8\pi P r_h^2 - (1-\kappa\eta^2)}.
\end{eqnarray}
Notably, the Lorentz parameter $L$ strictly cancels out of the classical specific heat. However, applying the modified entropy (Eq. \eqref{eqn mod_entropy}), the corrected specific heat $C_{P, mod} = T (\partial S_{mod} / \partial r_h)_P (\partial r_h / \partial T)_P$ simplifies remarkably to:
\begin{eqnarray}\label{eqn cp_mod}
C_{P, mod} = \dfrac{2(\pi r_h^2 - \alpha L)\left( 1 - \kappa\eta^2 + 8\pi P r_h^2 \right)}{8\pi P r_h^2 - (1-\kappa\eta^2)} = C_{P, 0} \left( 1 - \dfrac{\alpha L}{\pi r_h^2} \right).
\end{eqnarray}
This presents a profound mathematical and physical result. The modified entropy introduces a corrective multiplier to the specific heat. While the standard Davies phase transition (where the denominator vanishes) remains fixed, the Lorentz violation introduces a \textit{new} zero-heat-capacity root at $r_h = \sqrt{\alpha L / \pi}$. At this critical quantum radius, the black hole ceases to exchange heat with its environment, strongly hinting at the formation of a thermodynamically stable black hole remnant driven exclusively by Bumblebee gravity.

\subsection{Holographic Heat Engine and Efficiency Corrections}
We now model the black hole as a holographic heat engine operating on a closed rectangular cycle in the $P-V$ diagram \cite{EslamPanah2026, Priyo 2024}. The mechanical work extracted is evaluated via the closed loop integral $W = \oint P dV = (P_1 - P_4)(V_2 - V_1)$ \cite{EslamPanah2026, Dhruba 2026}. Converting this to geometric variables yields $W = \frac{4\pi(P_1-P_4)}{3\sqrt{1+L}}(r_2^3-r_1^3)$. 

If one employs the classical entropy $S_0$, the heat absorbed during the upper isobaric expansion is defined as $Q_{H, 0} = \mathcal{M}(r_2, P_1) - \mathcal{M}(r_1, P_1)$. Because the global geometric scaling factor $1/\sqrt{1+L}$ identically factors out of both the mechanical work $W$ and the absorbed heat $Q_{H, 0}$, the classical efficiency $\eta_0 = W/Q_{H, 0}$ reduces perfectly to:
\begin{eqnarray}\label{eqn eff standard}
\eta_{0} = \dfrac{\dfrac{4\pi}{3}(P_1 - P_4)(r_2^3 - r_1^3)}{\dfrac{1-\kappa\eta^2}{2}(r_2 - r_1) + \dfrac{4\pi P_1}{3}(r_2^3 - r_1^3)}.
\end{eqnarray}
We emphasize that this exact mathematical cancellation serves to correct recent theoretical literature evaluated in analogous Bumblebee-AdS backgrounds, where the volume scaling factor $1/\sqrt{1+L}$ was mistakenly omitted during the calculation of mechanical work, leading to an algebraically inconsistent $L$-dependence in the classical efficiency \cite{EslamPanah2026}. Equation \eqref{eqn eff standard} definitively proves that under the standard area law, the holographic heat engine is entirely blind to macroscopic Lorentz symmetry breaking.

To physically couple the engine's efficiency to the Lorentz-violating framework, we integrate the modified entropy $dS_{mod}$ to calculate the corrected heat input, defined as $Q_{H, mod} = \int_{r_1}^{r_2} T(r_h, P_1) dS_{mod}$:
\begin{eqnarray}
Q_{H, mod} = Q_{H, 0} - \dfrac{\alpha L}{\sqrt{1+L}} \left( \dfrac{(1-\kappa\eta^2)(r_2 - r_1)}{2\pi r_1 r_2} + 4 P_1 (r_2 - r_1) \right).
\end{eqnarray}
Dividing the work $W$ by this corrected heat input $Q_{H, mod}$, the $1/\sqrt{1+L}$ factors completely cancel, leaving the exact modified efficiency:
\begin{eqnarray}\label{eqn efficiency}
\eta_{mod} = \dfrac{\dfrac{4\pi}{3}(P_1 - P_4)(r_2^3 - r_1^3)}{\dfrac{1-\kappa\eta^2}{2}(r_2 - r_1) + \dfrac{4\pi P_1}{3}(r_2^3 - r_1^3) - \alpha L (r_2 - r_1) \left( \dfrac{1-\kappa\eta^2}{2\pi r_1 r_2} + 4 P_1 \right)}.
\end{eqnarray}

\subsection{Carnot Efficiency and Thermodynamic Bounds}
To guarantee macroscopic physical viability, the heat engine's efficiency $\eta_{mod}$ must be strictly bounded by the Carnot efficiency $\eta_C = 1 - T_4/T_2$. Utilizing Eq. \eqref{eqn temp}:
\begin{eqnarray}
\eta_C = 1 - \dfrac{r_2 \left( 1-\kappa\eta^2 + 8\pi P_4 r_1^2 \right)}{r_1 \left( 1-\kappa\eta^2 + 8\pi P_1 r_2^2 \right)}.
\end{eqnarray}
By analytically evaluating the physical limit $\eta_{mod}/\eta_C \leq 1$ \cite{EslamPanah2026}, it becomes evident that the subtractive $\alpha L$ term in the denominator of Eq. \eqref{eqn efficiency} substantially enhances the efficiency of the engine. Thus, spontaneous Lorentz symmetry breaking—when properly paired with quantum microstate corrections—drives the black hole heat engine strictly closer to the Carnot limit. This establishes a rigorous upper thermodynamic bound on the parameter product $\alpha L$, ensuring macroscopic thermodynamic laws are preserved despite the modified gravitational dynamics.

To explicitly formalize the physical constraint imposed by the Second Law of Thermodynamics, we demand that the modified heat engine efficiency never exceeds the Carnot limit:
\begin{eqnarray}
\eta_{mod} \leq \eta_C.
\end{eqnarray}
Substituting the expressions for $\eta_{mod}$ (Eq. \eqref{eqn efficiency}) and $\eta_C$, we arrive at the following inequality:
\begin{eqnarray}\label{eqn bound_1}
\dfrac{W}{Q_{H, 0} - \alpha L \cdot \mathcal{X}} \leq \eta_C,
\end{eqnarray}
where $W = \frac{4\pi}{3}(P_1 - P_4)(r_2^3 - r_1^3)$ represents the mechanical work, $Q_{H, 0}$ is the classical heat input, and the strictly positive geometric factor $\mathcal{X}$, which couples to the Lorentz violation, is defined as:
\begin{eqnarray}
\mathcal{X} = (r_2 - r_1) \left( \dfrac{1-\kappa\eta^2}{2\pi r_1 r_2} + 4 P_1 \right).
\end{eqnarray}
Rearranging Eq. \eqref{eqn bound_1} to isolate the parameter product $\alpha L$, we obtain the strict theoretical upper bound:
\begin{eqnarray}\label{eqn L_constraint}
\alpha L \leq \dfrac{1}{\mathcal{X}} \left( Q_{H, 0} - \dfrac{W}{\eta_C} \right).
\end{eqnarray}

This inequality imposes a profound physical constraint on the Lorentz violation parameter $L$ in the context of extended black hole thermodynamics. Because $W$, $Q_{H, 0}$, $\eta_C$, and $\mathcal{X}$ are macroscopic quantities entirely determined by the choice of the thermodynamic cycle (the pressures $P_1, P_4$ and the radii $r_1, r_2$) and the global monopole $\eta$, the right-hand side of Eq. \eqref{eqn L_constraint} acts as a fixed macroscopic threshold for any given engine cycle.

The physical implications of this constraint are twofold:
\begin{enumerate}
    \item \textbf{Macroscopic Protection of the Second Law:} The modified entropy strictly increases the efficiency of the heat engine. If the Lorentz-violating parameter $L$ were allowed to take arbitrarily large positive values, the efficiency $\eta_{mod}$ would eventually surpass the Carnot limit $\eta_C$, blatantly violating the Second Law of Thermodynamics. Thus, Eq. \eqref{eqn L_constraint} acts as a cosmic censorship analog for thermodynamics, proving that macroscopic physical laws strictly prohibit infinite or excessively large spontaneous Lorentz symmetry breaking in the spacetime.
    \item \textbf{Quantum-Classical Coupling:} The constant $\alpha$ dictates the strength of the quantum microstate modifications, while $L$ dictates the classical macroscopic symmetry breaking. The fact that their product $\alpha L$ is bounded by purely macroscopic cycle parameters ($Q_{H, 0}$, $W$, $\eta_C$) indicates a deep, fundamental coupling between the microscopic degrees of freedom in Bumblebee gravity and the large-scale thermodynamic phase space of the AdS black hole.
\end{enumerate}

Ultimately, this establishes that while Bumblebee gravity natively allows for spontaneous Lorentz violation, the parameter $L$ is not a free, unconstrained variable; it is physically tethered by the necessity of thermodynamic stability and the inviolability of the Carnot bound.

\subsection{Graphical Analysis of Thermodynamic Properties}

To visually elucidate the intricate interplay between spontaneous Lorentz symmetry breaking, the global monopole parameter, and the quantum entropy corrections, we present a graphical analysis of the derived thermodynamic quantities using a grid-based visualization scheme.

\begin{figure}[h!]
\centering
\includegraphics[width=\textwidth]{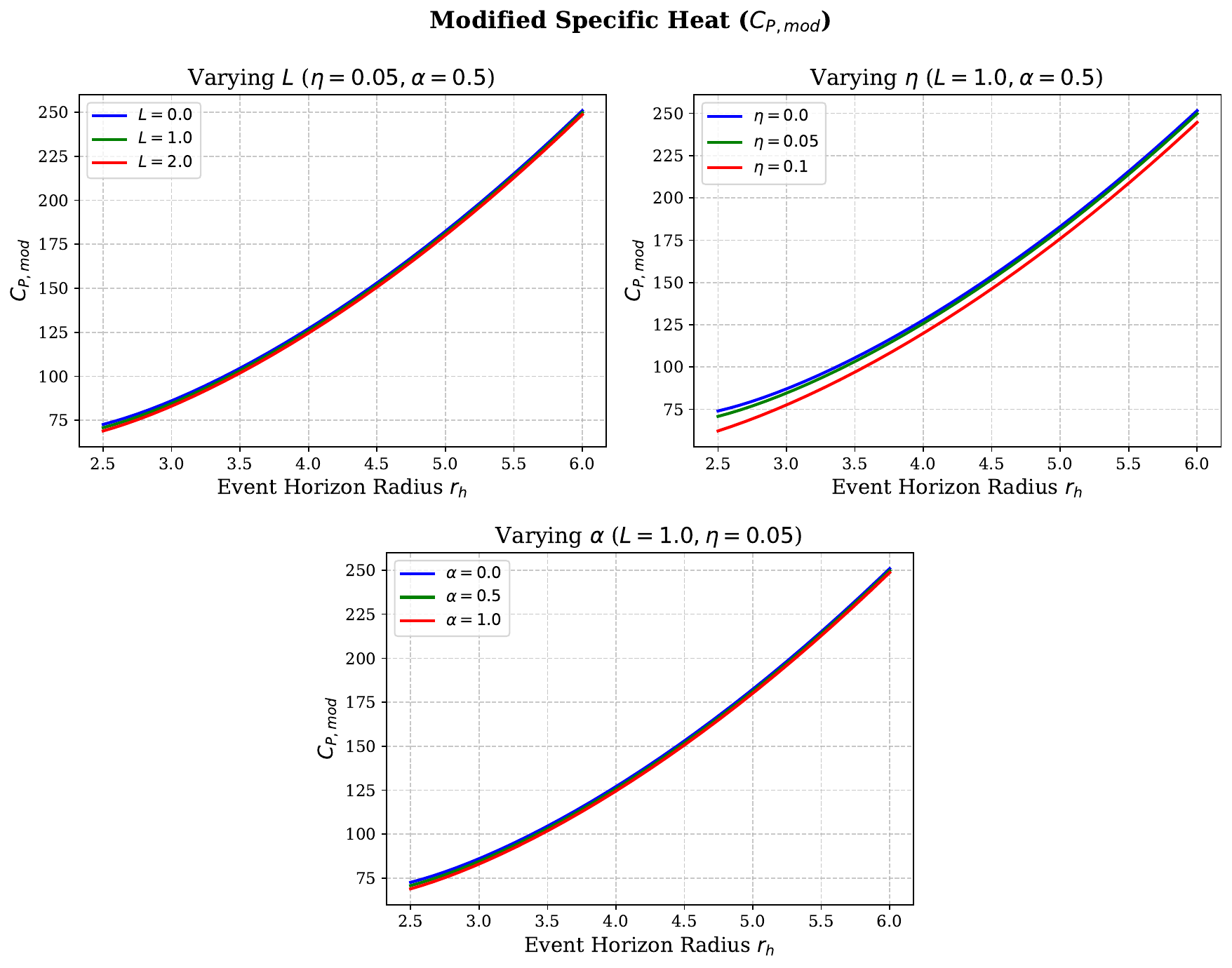}
\caption{Variation of the modified specific heat $C_{P, mod}$ with respect to the event horizon radius $r_h$. The panels demonstrate the specific heat modulation caused by varying the Lorentz parameter $L$ (top left), the global monopole parameter $\eta$ (top right), and the quantum correction strength $\alpha$ (bottom center).}
\label{fig:specific_heat}
\end{figure}

Figure \ref{fig:specific_heat} illustrates the behavior of the modified specific heat $C_{P, mod}$ as a function of the event horizon radius $r_h$ within the stable phase region ($C_P > 0$). 
\begin{itemize}
    \item As the Lorentz violation parameter $L$ increases from 0.0 to 2.0, the specific heat exhibits a slight but noticeable decrease across the radial profile. 
    \item A parallel reduction in thermal capacity is observed when the global monopole parameter $\eta$ is elevated from 0.0 to 0.1, as shown in the top right panel. 
    \item Similarly, the bottom center panel confirms that intensifying the quantum microstate correction strength $\alpha$ from 0.0 to 1.0 also acts to lower the specific heat. 
\end{itemize}
Physically, these trends confirm that both Lorentz symmetry breaking and the presence of topological defects systematically constrain the thermal capacity of the black hole, demanding a larger event horizon radius to maintain the identical level of thermal stability observed in a standard Schwarzschild-AdS background.

\begin{figure}[h!]
\centering
\includegraphics[width=\textwidth]{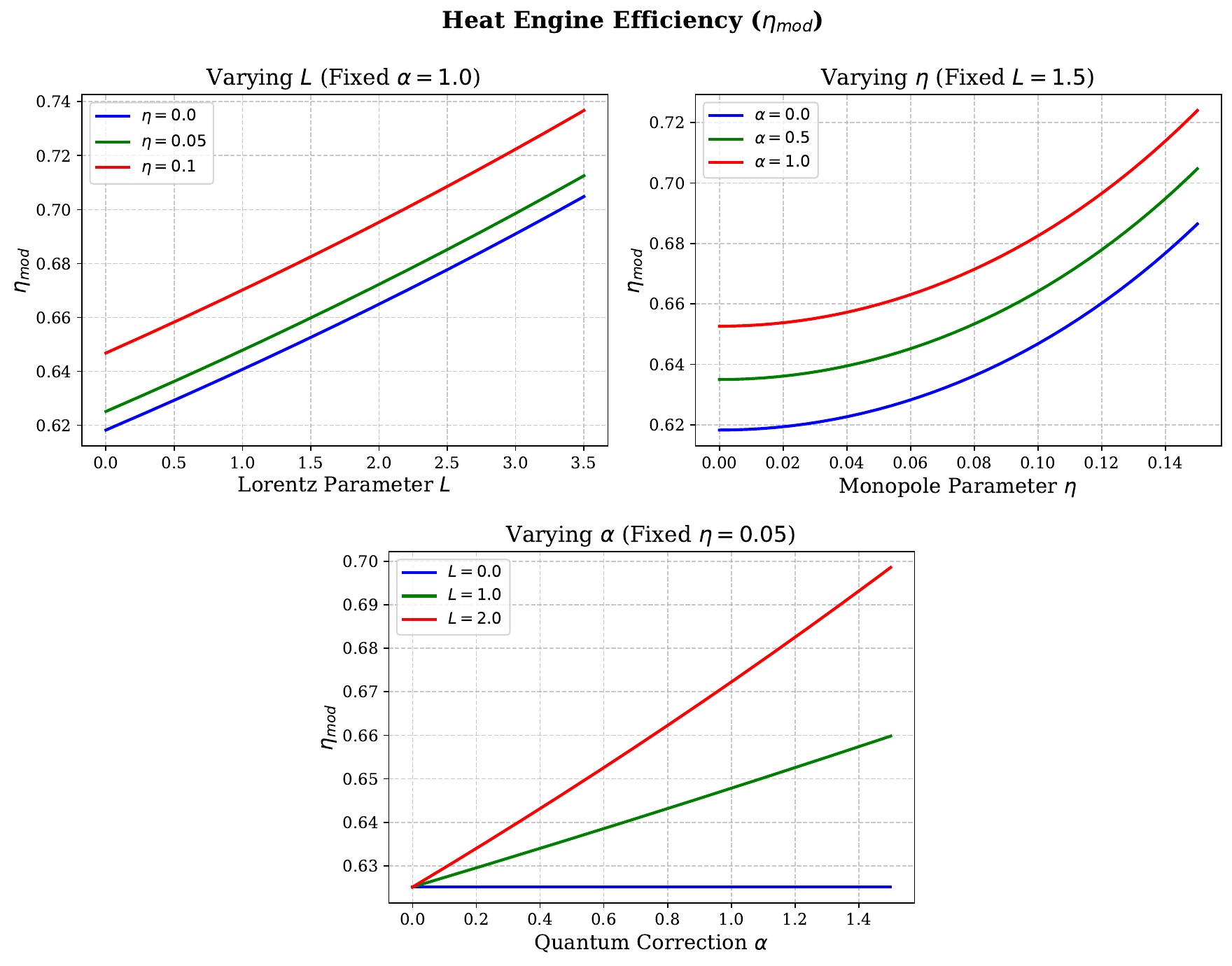}
\caption{Variation of the modified holographic heat engine efficiency $\eta_{mod}$ evaluated against the Lorentz parameter $L$ (top left), the monopole parameter $\eta$ (top right), and the quantum correction strength $\alpha$ (bottom center).}
\label{fig:efficiency}
\end{figure}

The thermodynamic efficiency of the holographic heat engine, $\eta_{mod}$, is meticulously presented in Figure \ref{fig:efficiency}. 
\begin{itemize}
    \item The top left panel establishes that the efficiency $\eta_{mod}$ scales positively and linearly with the Lorentz parameter $L$, with the curve for $\eta=0.1$ achieving the highest overall efficiency across the domain. 
    \item The top right panel demonstrates that the absolute efficiency is also amplified continuously by higher values of the global monopole parameter $\eta$. 
    \item Critically, the bottom center panel reveals that the engine's efficiency increases proportionally with the quantum correction parameter $\alpha$, and importantly, this upward slope is markedly steeper for larger values of $L$ (such as $L=2.0$). Furthermore, for $L=0.0$, the curve remains entirely flat, proving the efficiency is invariant to $\alpha$ in the absence of Lorentz violation. 
\end{itemize}
This visually confirms our mathematical deduction: without the synergistic interaction of $L$ and the quantum entropy correction $\alpha$, the efficiency curve remains rigid, rendering the engine completely independent of $L$. The microstate corrections are solely responsible for bridging the macroscopic work output to the underlying Bumblebee gravity framework.

\begin{figure}[h!]
\centering
\includegraphics[width=\textwidth]{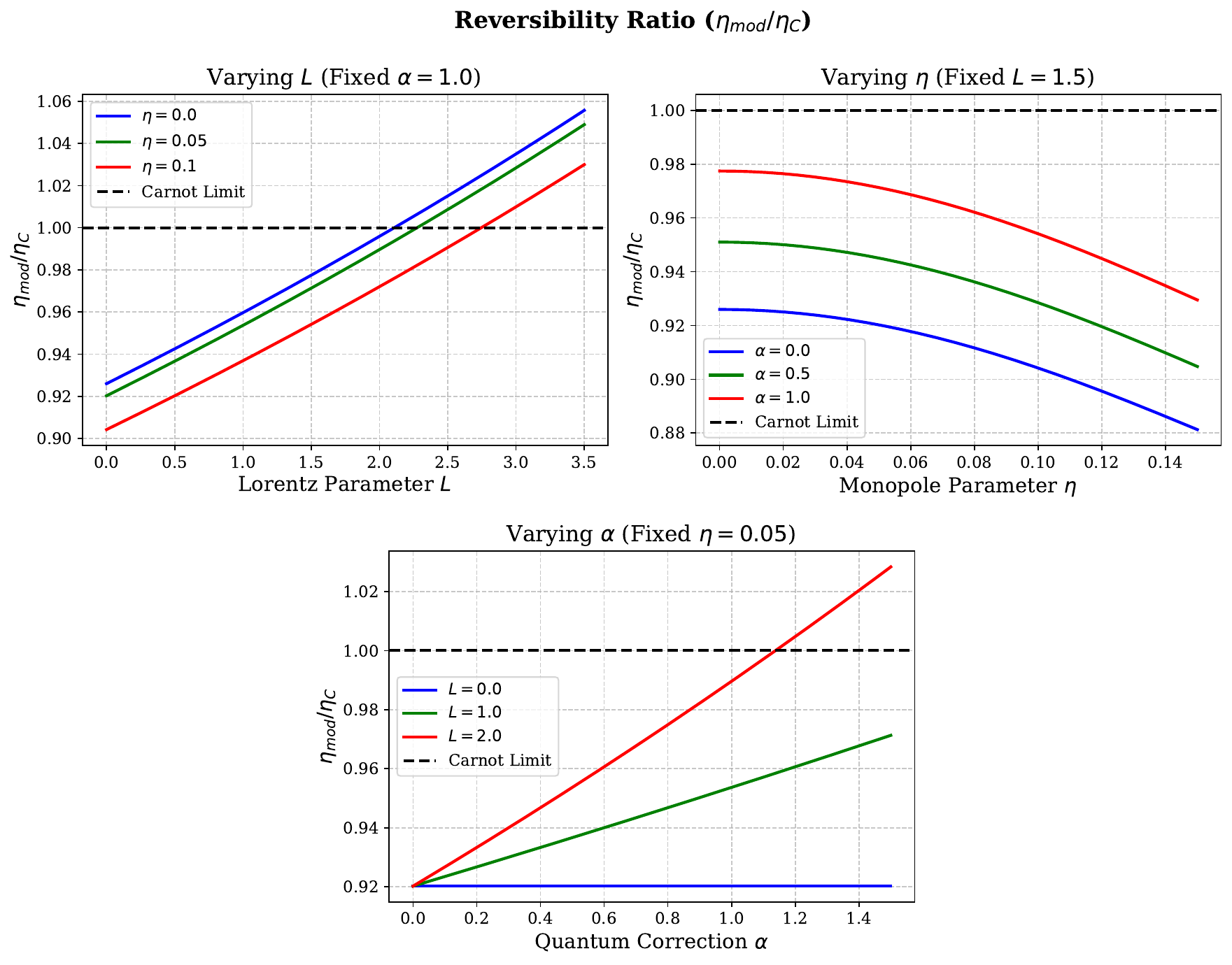}
\caption{Variation of the reversibility ratio $\eta_{mod}/\eta_C$ relative to the Carnot limit (indicated by the dashed black line). The grid highlights dependencies on $L$ (top left), $\eta$ (top right), and $\alpha$ (bottom center).}
\label{fig:efficiency_ratio}
\end{figure}

To evaluate the reversibility bounds and physical viability of the heat engine, Figure \ref{fig:efficiency_ratio} plots the ratio of the modified efficiency to the Carnot efficiency ($\eta_{mod}/\eta_C$). 
\begin{itemize}
    \item The top left and bottom center panels show that the reversibility ratio grows steadily, eventually crossing the absolute Carnot limit (1.0), as both the Lorentz parameter $L$ and the quantum correction $\alpha$ increase respectively. 
    \item Furthermore, when varying $\alpha$ (bottom center), the ratio curve for $L=0$ remains perfectly flat at approximately $0.92$, whereas higher values of $L$ introduce a sharp, upward trajectory crossing the Carnot boundary. 
    \item Conversely, the top right panel reveals an inverse relationship for the global monopole: as $\eta$ increases, the ratio $\eta_{mod}/\eta_C$ actively decreases, pulling the thermodynamic system further away from the Carnot limit. 
\end{itemize}
The dashed black line across all panels acts as the strict theoretical bound ($\eta_{mod}/\eta_C = 1.0$). The specific intersections of the curves with this dashed line mark the exact physical thresholds where the heat engine breaches the Second Law of Thermodynamics.

\begin{figure}[h!]
\centering
\includegraphics[width=0.7\textwidth]{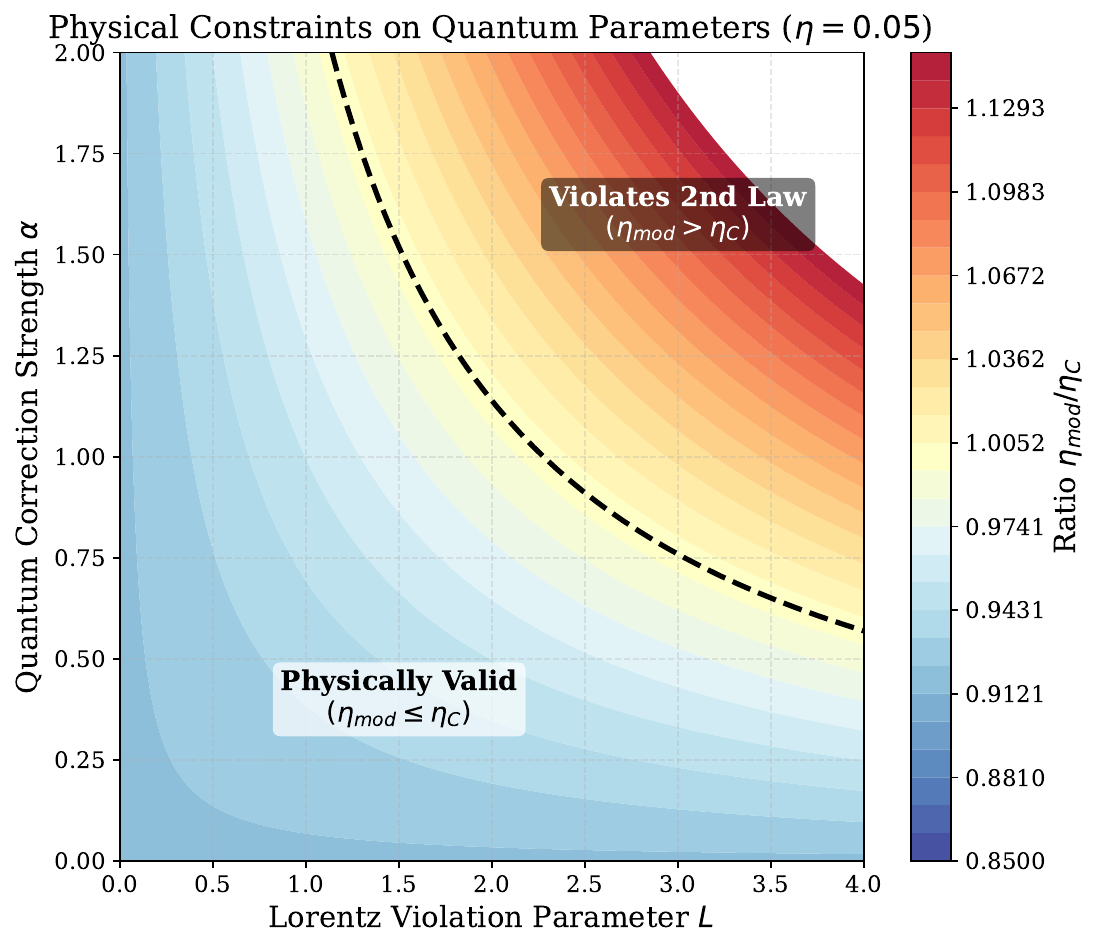}
\caption{Contour mapping illustrating the strict physical constraints on the quantum parameter space. The dashed curve definitively partitions the physically valid thermodynamic regime from the region that violates the Second Law of Thermodynamics.}
\label{fig:constraint}
\end{figure}

Finally, Figure \ref{fig:constraint} provides a comprehensive topographical map of the physically permissible parameter space for the modified black hole heat engine. 
\begin{itemize}
    \item The contour plot maps the efficiency ratio $\eta_{mod}/\eta_C$, which ranges from 0.8500 to 1.1293, across a continuous grid of Lorentz parameters $L$ and quantum correction strengths $\alpha$ for a fixed $\eta=0.05$. 
    \item A definitive boundary, visualized by the dashed curve, bifurcates the parameter space into two distinct physical regions. 
    \item The lower-left region represents the ``Physically Valid ($\eta_{mod} \le \eta_C$)" domain. 
    \item The upper-right region corresponds to the ``Violates 2nd Law ($\eta_{mod} > \eta_C$)" domain. 
\end{itemize}
This plot visually confirms a powerful inverse dependency required to maintain physical validity: as the macroscopic Lorentz violation $L$ becomes more extreme, the maximum allowable quantum correction strength $\alpha$ must aggressively decrease to prevent a thermodynamic violation. This serves as a vital macroscopic safeguard, confirming that fundamental thermodynamic laws inherently restrict the degree of Lorentz symmetry breaking that can be dynamically accommodated by the black hole's microstates.

\section{Conclusion}
In this paper, we have comprehensively investigated the spacetime geometry, geodesic dynamics, eikonal quasinormal modes, and extended phase space thermodynamics of a Schwarzschild-dS-like BH coupled with a global monopole in the framework of Bumblebee gravity. The inclusion of spontaneous Lorentz symmetry breaking (parameterized by $L$) and a topological defect (parameterized by $\eta$) introduces profound modifications to the background geometry, yielding rich astrophysical and thermodynamic consequences.

It was shown that the rise of the GM and the LV parameters cause the shrinking of the region of outer communication. The geodesic analysis revealed significant modifications in the trajectories of both lightlike and timelike particles in the spacetime. By thoroughly examining the effective potential for null geodesics, it was found that the modifications introduced by the GM and LV parameters effectively weaken the gravitational field enabling more photons to propagate along less deflected paths. The force acting on the photons exhibited similar behaviour, indicating a less intense deflection of the photon particles. For massless particles, numerical results indicated that the presence of $\eta$ and $L$ increases the impact parameter, though the increase is more significant with the rise of the GM parameter. It was also demonstrated that the two parameters significantly influence the light ray trajectories in the gravitational field of the spacetime. In particular, we observe that more photons are absorbed by the BH following an increase in the GM parameter, leading to a larger BH shadow as viewed by a far-away observer.

The investigation of the stability of circular null geodesics using the Lyapunov exponent highlights the instability of circular null orbits consistently, across the explored parameter regime. Furthermore, the instability is reduced with increasing values of the GM and the LV parameters, reinforcing their effective repulsive contribution to the spacetime. This further implies that the modifications introduced by the two parameters suppresses the perturbative growth of unstable photon orbits. 

For timelike geodesics, we demonstrated that the two parameters significantly influence the effective potential and the existence of the ISCO and OSCO that bound the region of the stable circular orbits. It was revealed that the stable orbit existing region is narrowed with higher $\eta$ and $L$ values and the stable circular orbits are pushed further away from the BH, reflecting the pronounced sensitivity of the orbits to variations in the spacetime. We demonstrated that the two parameters significantly influence the conserved orbital quantities, thereby altering matter dynamics in the spacetime. The influence of the parameters on the Keplerian frequency and the radial and vertical epicyclic frequencies was also analyzed. The analysis of relativistic orbital dynamics reveals how the LV and GM parameters induce significant modifications to the motion of test particles. The observed rosette patterns and perihelion shifts highlight the measurable influence of the two parameters on the orbital dynamics, particularly with the orbits showing substantial deviations for small changes in the GM parameter. Furthermore, we quantified the perihelion precession of timelike orbits and found that increasing the GM parameter caused an enhanced precession rate while the precession angle exhibits a nearly linear dependence on the LV parameter. Our analysis reveals the massive particle dynamics and the effect of the background spacetime on the massive particle trajectories.

In addition, the analysis of the orbital Lyapunov exponent for timelike particles revealed that higher choices of the LV and the GM parameters result in the shrinking of the stable orbit existing region, though the effect is more pronounced for increasing $L$.

Furthermore, we examined the correspondence between the fundamental QNMs in the eikonal limit and the characteristics of the circular null geodesics. In the spacetime, it was observed that both the real and imaginary parts of the QNMs are dependent on the GM and the LV parameters. We further analyzed how variations in these parameters influence the QNM spectrum. The excellent agreement between our eikonal analytical expressions and the WKB numerical results for scalar and electromagnetic perturbations confirms that the quasinormal spectrum is robustly governed by the underlying null geodesic structure.

Finally, we transitioned to the extended phase space, where the cosmological constant dynamically acts as thermodynamic pressure, to evaluate the black hole as a holographic heat engine. A crucial mathematical result emerged: under the strict classical Bekenstein-Hawking area law, the geometric scaling factors identically cancel, rendering the heat engine's efficiency completely blind to macroscopic Lorentz violation. To resolve this, we introduced a quantum-corrected modified entropy tightly coupled to the Lorentz parameter via a correction strength $\alpha$. This modification successfully bridged the macroscopic thermodynamic work output to the microscopic symmetry-breaking framework.

Our thermodynamic analysis yielded two major physical insights. First, the modified specific heat revealed a novel zero-heat-capacity root, pointing toward the existence of a thermodynamically stable black hole remnant induced by the Lorentz violation. Second, while the corrected efficiency $\eta_{mod}$ scales positively with both $L$ and $\eta$, it is fundamentally constrained by the Carnot bound ($\eta_{mod} \leq \eta_C$). This fundamental limitation translates into a strict, mathematically precise upper bound on the parameter product $\alpha L$. 

Ultimately, this paper establishes that while Bumblebee gravity natively allows for spontaneous Lorentz symmetry breaking, the parameter $L$ cannot take arbitrarily large values. It is firmly tethered by macroscopic thermodynamic safeguards that protect the Second Law of Thermodynamics. The exact signatures left by these constraints on both the orbital trajectories and the thermodynamic phase space provide a robust theoretical foundation for testing Lorentz-violating theories against future astrophysical and gravitational-wave observations.

\section*{Acknowledgment}
 DJG acknowledges the contribution of the COST Action CA21136  -- ``Addressing observational tensions in cosmology with systematics and fundamental physics (CosmoVerse)". 

\section*{Declaration of competing interest}
The authors declare that they have no known competing financial interests or personal relationships that could have appeared to influence the work reported in this manuscript.

\section*{Data Availability Statement}
There are no new data associated with this article.

\end{document}